# Optimized Analysis of the AC Magnetic Susceptibility Data in Several Spin-Glass Systems using the Vogel-Fulcher and Power Laws

Mouli Roy-Chowdhury[1], Mohindar S. Seehra[2] and Subhash Thota[1,*]

[1]*Department of Physics, Indian Institute of Technology Guwahati, Guwahati -781039, India*
[2]*Department of Physics and Astronomy, West Virginia University, Morgantown, West Virginia 26506, USA*

## Abstract

In spin-glasses (SG), the relaxation time $\tau$ $(= 1/2\pi f)$ vs. $T_f$ data at the peak position $T_f$ in the temperature variation of the *ac* magnetic susceptibilities at different frequencies $f$ is often fit to the Vogel-Fulcher Law (VFL): $\tau = \tau_0 \exp[E_a/k_B(T_f - T_0)]$ and to the Power Law (PL): $\tau = \tau_0^* [(T_f - T_{SG})/T_{SG}]^{-zv}$. Both these laws have three fitting parameters each, leaving a degree of uncertainty since the magnitudes of the evaluated parameters $\tau_0$, $E_a/k_B$, $\tau_0^*$ and $zv$ depend strongly on the choice of $T_0$ and $T_{SG}$. Here we report an optimized procedure for the analysis of $\tau$ vs. $T_f$ data on several SG systems for which we could extract such data from published sources. In this optimized method, the data of $\tau$ vs. $T_f$ are fit by varying $T_0$ in the linear plots of $Ln \tau$ vs $1/(T_f - T_0)$ for the VFL and by varying $T_{SG}$ in the linear plot of $Ln \tau$ vs. $\ln (T_f - T_{SG})/T_{SG}$ for the PL till optimum fits are obtained. The analysis of the associated magnitudes of $\tau_0$, $E_a/k_B$, $\tau_0^*$ and $zv$ for these optimum values of $T_0$ and $T_{SG}$ shows that magnitudes of $\tau_0^*$, $\tau_0$ and $zv$ fail to provide a clear distinction between canonical and cluster SG. However, new results emerge showing $E_a/(k_B T_0) < 1$ in canonical SG whereas $E_a/(k_B T_0) > 1$ for cluster SG systems and the optimized $T_0 <$ optimized $T_{SG}$ in all cases. Although some interpretation of these new results is presented, a more rigorous theoretical justification of the boundary near $E_a/(k_B T_0) \sim 1$ is desired along with testing of these criteria in other SG systems.

*subhasht@iitg.ac.in

## 1. Introduction

The spin-glass (SG) phase has enamoured physicists since its discovery in the 1970s with Edwards and Anderson proposing the first basic model in 1975 [1]. Giorgio Parisi winning the Nobel Prize in Physics in 2021 for "the discovery of the interplay of disorder and fluctuations in physical systems from atomic to planetary scales" has made the SG problem as relevant as ever [2]. Apart from the original problem of random magnets, it now serves to answer a wide-range of questions catering to various fields of science from computer science to economy to biology [3-6].

The SG phase lacks long-range magnetic order, analogous to structural glasses which lack long-range structural order [7] and hence, the assigned SG nomenclature. The earlier studies on SG systems are summarized in the review by Binder and Young [8] and in the book by Mydosh [9]. The two well-known classes of SG are: (i) canonical SG consisting of individual magnetic impurities present in a non-magnetic system with long-range inter-spin interactions such as AuMn containing 2.98% Mn; and (ii) cluster SG usually consisting of magnetic ions with short-range inter-spin interactions and larger concentrations of magnetic ions but below the percolation threshold. Generally, the SG phase is characterised by multiple equilibrium ground states like free-energy valleys separated by energy barriers. Hence the SG phase constitutes a new state of magnetism different from the long-range ordered (LRO) magnetic states like ferromagnets, ferrimagnets, and antiferromagnets. However, in SG systems, the collective freezing of spins does occur below a freezing temperature $T_{SG}$ but without LRO, typically resulting from random site occupancy and geometric frustration along with competing exchange interactions.

Experimentally, the SG phase shows some distinct features: (*i*) a frequency-dependent shift of the peak temperature $T_f$ in the temperature dependence of the *ac* magnetic susceptibility ($\chi'$ and $\chi''$) data; (*ii*) absence of an anomaly or at best a broad maximum around $T_{SG}$ in the specific heat $C_p$ vs $T$ data instead of a lambda type asymmetric peak observed for LRO systems; (*iii*) non-exponential time dependence of magnetization, and (*iv*) memory effects [8,9]. The spin relaxation in SG systems is due to the multivalley ground state, these valleys being separated by temperature-dependent energy barriers of different magnitudes [8,9]. On cooling the system through $T_{SG}$, the system lands in one of these valleys from which it can relax by overcoming these barriers. This makes the relaxation in SG dependent on external perturbations such as applied magnetic field and *ac* frequency. The bifurcation of the field-cooled and zero-field cooled magnetization below $T_{SG}$, shift of $T_{SG}$ with applied H given by the de Almeida –Thouless [10] or Gabay-Toulouse [11] lines, non-linear susceptibility, magnetic viscosity, aging phenomenon, and the frequency-dependent shift of the peak temperature $T_f$ in the *ac* susceptibility are some of the techniques used for investigating spin relaxation in SG systems [8,9,10,11].

In this work, we have focussed on the analysis of $T_f$ vs. *f* data obtained from *ac* susceptibility studies in seventeen (17) SG systems for which we could extract the data from published sources. Temperature and frequency dependence of the *ac* susceptibilities, χ' and χ'', are often used to distinguish between magnetic nanoparticles and canonical vs. cluster SG systems. For this purpose, the temperature shift in the peak position $T_f$ in χ'' or χ' with change in frequency *f* is used to define the Mydosh parameter Ω [8,9,12,13]:

$$\Omega = (T_{f2} - T_{f1})/T_{f1}(log f_2 - log f_1) \qquad (1)$$

where $f_1$ and $f_2$ are two sufficiently different frequencies. For magnetic nanoparticles, it is generally observed that $\Omega > 0.05$ with the magnitude of $\Omega$ increasing with decrease in the interparticle interaction [13]. For $0.01< \Omega < 0.05$, the system is usually classified as a cluster SG, whereas for a canonical SG system, $\Omega \sim 0.005$ is an order of magnitude smaller [8,9,12,13]. For 3D magnetic transitions with long-range magnetic ordering, $\Omega \approx 0$ is often reported [14,15].

Additional information on the spin dynamics of such systems is obtained from the fit of the relaxation time τ vs. $T_f$ data to the Vogel-Fulcher law (VFL) given by [8,9,16]:

$$\tau = \tau_0 \exp[E_a/k_B(T_f - T_0)] \qquad (2)$$

and the Power Law (PL) given by [8,9,17-20]:

$$\tau = \tau_0^*[(T_f - T_{SG})/T_{SG}]^{-zv} \qquad (3)$$

For use in Eqs. (2) and (3), the relaxation time τ =1/2π*f* is determined from the frequency *f* whereas $T_f$ represents the temperature of the peak positions in the *ac* susceptibilities χ' (or χ'') vs. *T* data at different *f* thus yielding the τ vs. $T_f$ data. Other parameters in Eqs. (2) and (3) are: activation energy $E_a$, *zv*, $\tau_0$, and $\tau_0^*$, the relaxation times of individual spins or clusters; $T_0$, the strength of interparticle or inter-cluster interactions; $T_{SG}$, the SG temperature; and *zv*, the dynamical critical exponent. In general, for most reported SG systems, *zv* lies between 4 and 12 and the magnitude of $\tau_0^*$ usually falls between $10^{-12}$ and $10^{-13}$s for canonical SG systems, whereas, in general, $\tau_0{}^*$ for the cluster SG is significantly higher and lies in the range $10^{-7}$ to $10^{-10}$ s [8,9,19,21-26]. However, the consensus about the range of values for these parameters is often contradictory [27-30] and how the magnitudes of $T_{SG}$ and $T_0$ are selected is often not explained.

The determination of $\Omega$ from the data using Eq. (1) is straightforward. However, fitting the $\tau$ vs. $T_f$ data to Eqs. (2) and (3) is possible for a range of these fitting parameters because both these laws have three fitting parameters each, thus leaving a degree of uncertainty. Specifically, the magnitudes of the evaluated parameters $\tau_0$, $E_a/k_B$, $\tau_0{}^*$ and $z\upsilon$ depend on the choice of $T_0$ and $T_{SG}$. As noted in our recent paper on the spinel ZnTiCoO$_4$ (listed as ZTCO hereafter) [31], the linear plots of $Ln\ \tau$ $vs$ $1/(T_f - T_o)$ for different choices of $T_0$ for the VFL and $Ln\ \tau$ vs. $Ln\ (T_f - T_{SG})/T_{SG}$ for different choices of $T_{SG}$ for the PL are possible. These linear fits yield $\tau_0$ and $E_a/k_B$ for each choice of $T_0$ for the VFL and $\tau_0{}^*$ and $z\upsilon$ for each choice of $T_{SG}$. Using different values of $T_{SG}$ and $T_0$ still yielded respectable linear fits to Eqs. (2) and (3), yet magnitudes of the fitting parameters were quite different for each choice of $T_{SG}$ and $T_0$. So, a procedure was developed by plotting the variations of the evaluated parameters and associated adjusted R$^2$ (AR$^2$) value of the linear fits against $T_0$ and $T_{SG}$ [32]. The quality of the linear fits was determined from the maximum value of AR$^2$, with AR$^2$ =1 being valid for a perfect linear fit [32]. The analysis of the data for ZTCO yielded the optimum (maximum) value of AR$^2$ = 0.993 for a particular $T_{SG}$ and $T_0$ [31]. The corresponding magnitudes of $\tau_0$ and $E_a$ in Eq. (2) and $\tau_0$* and $z\upsilon$ in Eq. (3) were then considered as optimum magnitudes of these parameters for the system.

In this paper, we have employed this procedure to reanalyse the data of $\tau$ vs. $T_f$ in seventeen (17) randomly chosen SG systems for which we could extract such data from published sources. Results of the evaluated parameters $\tau_0$ and $E_a/k_B$ in Eq. (2) and $\tau_0$* and $z\upsilon$ in Eq. (3) from this optimized analysis for the seventeen SG systems along with the evaluated magnitudes of $\Omega$ are collected in Table I with details of the analysis given in the following pages. This analysis shows that magnitudes of $\tau_0$* and $z\upsilon$ alone are inadequate to distinguish between different classes of SG. Instead, it is proposed that the magnitudes of $\Omega$ along with that of the ratio $E_a/(k_B T_o)$ can be used to distinguish between the canonical and cluster SG unambiguously. Typically, $\Omega \sim 10^{-3}$ for canonical SG while $\Omega \sim 10^{-2}$ in cluster SG as also previously reported in literature and used by many investigators to distinguish between canonical and cluster SG as also discussed below. The new results reported below from the analysis presented in this work are that the ratio $E_a/k_BT_0 < 1$ in canonical SG whereas $E_a/k_BT_0 > 1$ for cluster SG, and the optimized $T_0 <$ optimized $T_{SG}$. Some interpretations of these new results are also presented along with suggestions for future studies.

## 2. Examples of various spin-glasses

Analysis of the $\tau$ vs. $T_f$ data in seventeen (17) SG systems is presented here using the peak temperature $T_f$ of the *ac* susceptibilities and $\tau = 1/2\pi f$ for each frequency $f$. For these systems, we could extract the data of $T_f$ vs $f$ from the *ac* susceptibilities from published sources using the software, WebPlotDigitizer-4 [33]. But for reasons of brevity, we show the details of the data and analysis for only two systems whereas corresponding figures for the other 15 systems are given in the SI (supplemental Information [34]). Typically, these measurements were done with *ac* magnetic field with amplitude ∼ 5 Oe.

### 2.1. Canonical spin-glasses:

**2.1.1. AuMn:** In Fig. 1, we show the replotted data of Mulder *et al* [35] on the temperature dependence of $\chi'$ vs T at different frequencies for a well-known canonical SG system, AuMn$_{2.98\%}$ containing 2.98% Mn, who reported $T_{SG} = 10.23$ K and $\Omega = 0.0045$. We re-analyzed the $\tau$ vs. $T_f$ data obtained from Fig.1 using the optimized AR$^2$ method in which plots of *Ln* $\tau$ *vs* $1/(T_f - T_0)$ for the VFL and *Ln* $\tau$ vs. *Ln* $(T_f - T_{SG})/T_{SG}$ for the PL for different choices of $T_0$ and $T_{SG}$ were made with the linear fits yielding $\tau_0$ and $E_a/k_B$ for each choice of $T_0$ in case of the VFL and $\tau_0$* and *zv* for each choice of $T_{SG}$ for the PL. These plots depicting the variations of the evaluated parameters for AuMn with change in $T_0$ and $T_{SG}$ are shown in Fig 2. Here, the optimum $T_{SG} = 10.0$ K with maximum AR$^2 = 0.997$ is observed for the PL analysis with the corresponding optimum $\tau_0$* $= 3.6 \times 10^{-16}$ s, and *zv* = 7.6 whereas in the VFL analysis, peak with AR$^2 = 0.997$ occurs at $T_0 = 9.7$ K with $\tau_0 = 1.7 \times 10^{-11}$ s and $E_a/k_BT_0 = 0.98$. Our analysis also yielded $\Omega = 0.006$ for AuMn. According to our suggested classification ($E_a/k_BT_0 < 1$ for canonical SG and $E_a/k_BT_0 > 1$ for cluster SG), AuMn is identified as a canonical SG.

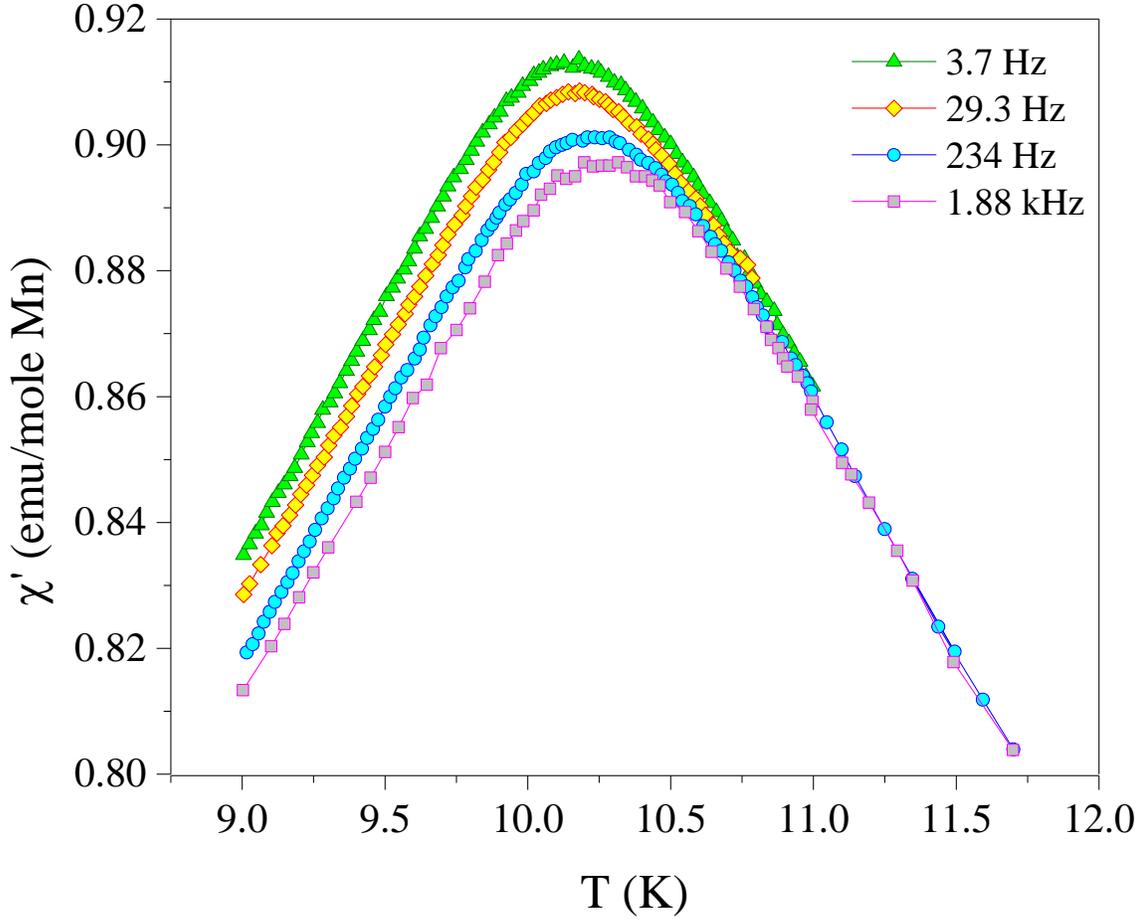

**Figure 1:** Frequency dependence of the real part of *ac* susceptibility χ′ versus temperature extracted from Figure 2 of Mulder *et al* [35] around the freezing temperature of AuMn$_{2.98\%}$ using the software WebPlotDigitizer-4 [33].

The results in Fig.2 show that with increase in $T_{SG}$, magnitude of $\tau_0^*$ increases rapidly whereas $zv$ decreases almost linearly. Similarly, with increase in $T_0$, $\tau_0$ increases rapidly whereas $E_a/k_B$ decreases linearly. This is a common theme for all the results presented here, necessitating the need for determining the optimum values of $T_{SG}$ and $T_0$ as reported in this work.

**2.1.2. CuMn:** The data of $T_f$ vs. *f* for the CuMn$_{4.6\%}$ system was reported in 1980 by Tholence whose initial analysis of this data assuming $\tau_0 = 10^{-13}$s for the VFL yielded $T_0 = 25.5$K and $E_a/k_B = 59$ K [36]. However, a more detailed analysis by Souletie and Tholence in 1985 [37] reported the following fitting parameters for this system: $T_{SG} = 27.45$ K with $\tau_0^* = 7.7 \times 10^{-13}$ s, and $zv = 5.5$ for the PL and $T_0 = 26.9$ K with $\tau_0 = 4.0 \times 10^{-8}$ s and $E_a/k_B = 11.75$ K for the VFL. Our analysis of this data with maximum AR$^2$ = 0.9975 (shown in Fig. S1) yielded the following optimum parameters: $T_{SG}$ = 27.3 K, $\tau_0^*$ = 4.8×10$^{-15}$ s, and $zv$ = 6.82 for the PL and $T_0$ = 26.7 K with $\tau_0$ = 8.8×10$^{-10}$ s and $E_a/k_B$ = 16.7 K for the VFL. These optimum values from our analysis, also listed in Table I, are only slightly different from those reported in [37]. With $\Omega$ = 0.0084 and $E_a/k_B T_0$ = 16.7/26.7 = 0.625 < 1, this system is classified as a canonical SG.

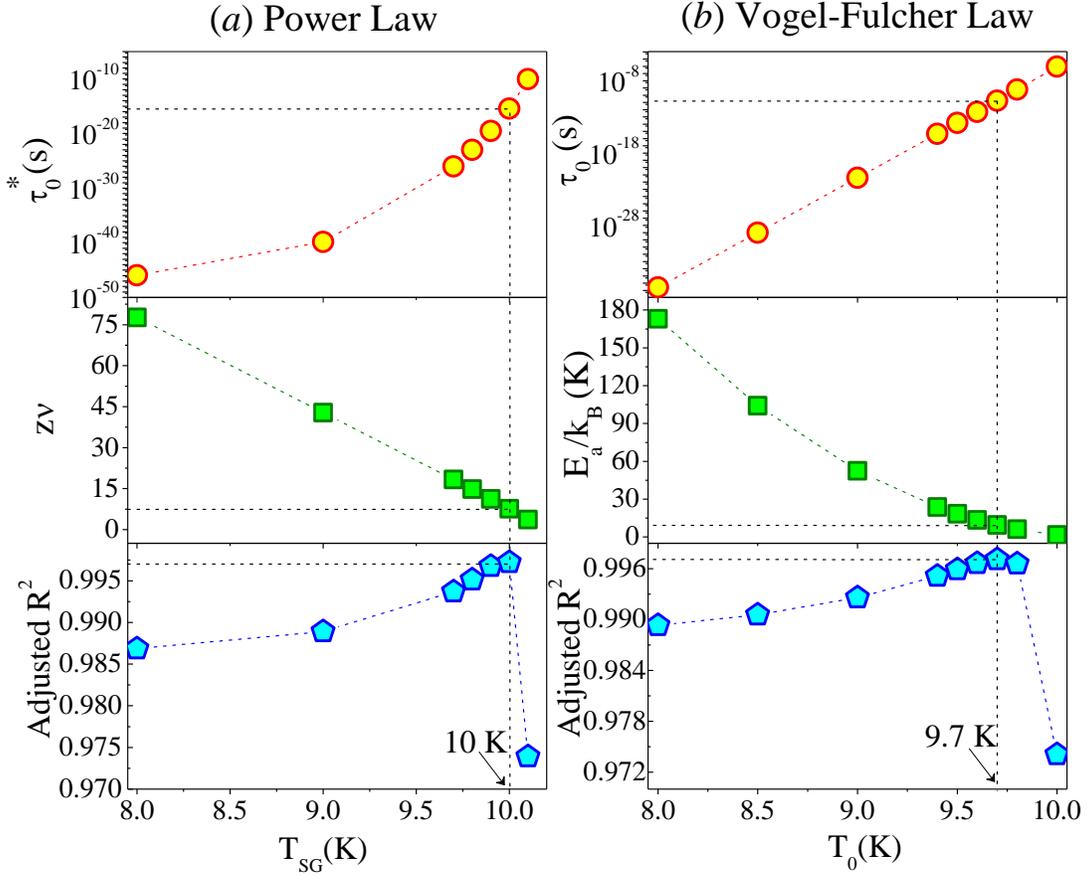

(a) Power Law  (b) Vogel-Fulcher Law

**Figure 2**: Plotted are the variations of the evaluated parameters for AuMn$_{2.98\%}$ as a function of different choices of $T_{SG}$ in the Power Law (Eq. 3) and as a function of $T_0$ in the Vogel-Fulcher Law (Eq. 2) with the dotted lines connecting the data points as visual guides. The vertical dotted lines mark the positions of the maximum AR$^2$ and the corresponding optimum magnitudes of $\tau_0^*$, $zv$ and $T_{SG}$ in the Power Law and optimum magnitudes of $T_0$, $\tau_0$ and $E_a/k_B$ in the Vogel-Fulcher Law. See Table I for listing of the optimum parameters for different systems.

### 2.1.3. Na$_{0.7}$MnO$_2$:

Luo *et al* reported experimental studies on the hexagonal SG candidate Na$_{0.70}$MnO$_2$ including the frequency-dependent *ac* magnetic susceptibility measurements [38] with $T_{SG}$ = 39.0 K and $\Omega$ = 0.004, the latter falling in the range observed in canonical SG. By extracting the data of χ' vs. T at different $f$ from [38], we carried out our analysis like the one shown in Fig. 2 for AuMn, see Fig. S2 in SI [34]. Results from our analysis yielded $\Omega = 0.005$, optimum $T_{SG} = 38.6$ K, $\tau_0* = 1.5 \times 10^{-14}$ s, and $zv = 6.0$, optimum $T_0 = 38.1$ K, $\tau_0 = 1.3 \times 10^{-8}$ s, and $E_a/k_B = 12.3$ K, the latter yielding $E_a/k_B T_0 = 0.32 < 1$. Using our criteria, $Na_{0.70}MnO_2$ is a canonical SG. This agrees with the conclusion of [38] which was based just on the magnitude of $\Omega = 0.004$ since analysis of the data using the PL and VFL was not reported.

**2.1.4. $CaSrFeRuO_6$:** From their magnetic studies in the atomically disordered perovskite $CaSrFeRuO_6$, including the fitting of the frequency-dependent *ac* susceptibility data to conventional scaling laws given by Eqs. (2) and (3), Naveen et al [39] reported a SG transition with the following parameters: $T_{SG} = 62$ K, $\tau_0* = 7.57 \times 10^{-12}$ s, $zv = 5.4$, $T_0 = 58$ K, $\tau_0 = 1.85 \times 10^{-12}$ s and $E_a/k_B = 119$ K, the latter yielding $E_a/k_B T_0 = 2.05$. The nature of the transition was suggested to be intermediate between canonical and cluster SG [39]. The results from our optimized analysis of their data (figure S3 [34]), yielded $\Omega = 0.009$, optimum $T_{SG} = 63.6$ K, $\tau_0^* = 6.2 \times 10^{-9}$ s, $zv = 2.0$, optimum $T_0 = 63.2$ K, $\tau_0 = 2.1 \times 10^{-6}$ s and $E_a/k_B = 3.3$ K, leading to $E_a/k_B T_0 = 0.05$ (see Table I). It is evident that the magnitudes of these parameters from our analysis differ significantly from those reported by Naveen *et al*, particularly the ratio $E_a/k_B T_0$. Based on $\Omega = 0.009$ and $E_a/k_B T_0 = 0.05 < 1$ obtained in our analysis, we classify this system as a canonical SG.

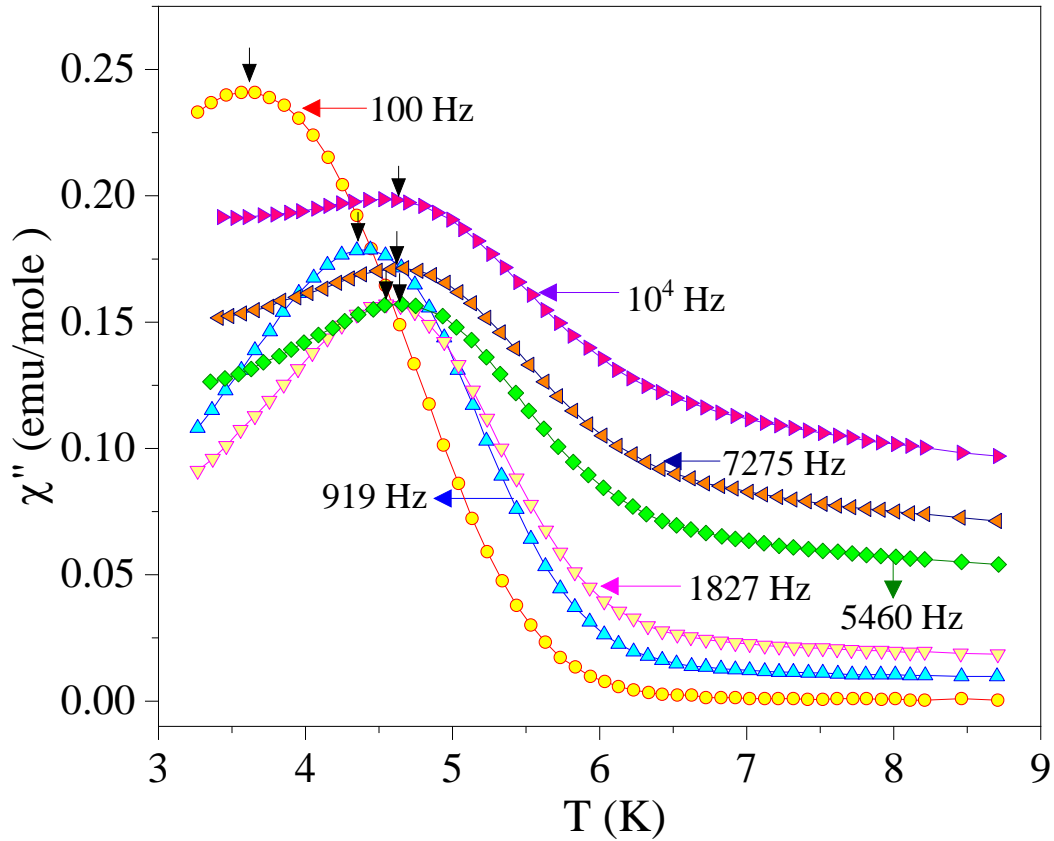

**Figure 3:** Temperature dependence of the ac-magnetic susceptibility χ'' measured at different frequencies extracted from Figure 2(b) of Anand *et al* [19] around the spin-freezing temperature of PrRhSn₃ using the software WebPlotDigitizer-4 [33]. The black arrows mark the peak positions for the χ'' versus T data at different frequencies.

**2.1.5. LiFeSnO₄-HT:** The analysis of the variation of $T_f$ with $f$ for the high-temperature (HT) phase of LiFeSnO₄= $Li^+Fe^{3+}Sn^{+4}O_4$ was reported by Banerjee *et al* [40] without listing $T_{SG}$. For this phase with orthorhombic structure, they reported $\Omega = 0.015$, $\tau_0^* = 3 \times 10^{-13}$ s, $zv = 7.3$, $T_0 = 18.7$ K, $\tau_0 \sim 10^{-10}$ s and $E_a/k_B = 67.5$ K yielding $E_a/k_B T_0 = 3.6$. They classified it as a canonical SG based primarily on $\tau_0^* = 3 \times 10^{-13}$ s. Our analysis of the data extracted from [40] is shown in Fig. S4 [34] leading to optimum $T_{SG} = 21.3$ K, $\tau_0^* = 3.6 \times 10^{-9}$ s, $zv = 2.8$, $T_0 = 20.8$ K, $\tau_0 = 3.6 \times 10^{-7}$ s and $E_a/k_B = 5.7$ K. Since $E_a/k_B T_0 = 0.27 < 1$, it is a canonical SG according to our criteria which agrees with the conclusion in [40]. However, note the substantially different magnitudes of the parameters resulting from our analysis, especially the magnitude of $E_a/(k_B T_0)$.

**2.1.6. IrMnGa**: In a recent work, Kroder *et al* [41] reported IrMnGa as a canonical SG, based primarily on the magnitude of $\Omega = 0.01$. They also reported $T_{SG} = 72.0$ K, $\tau_0^* = 4 \times 10^{-12}$ s, $zv = 6.2$, $T_0 = 67.4$ K, $\tau_0 \sim 10^{-11}$ s, and $E_a/k_B = (141 \pm 42)$ K, the latter yielding $E_a/k_B T_0 = 2.1 \pm 0.6$. Our optimized analysis of their $T_f$ vs $f$ data (see Fig. S5 in [34]) yielded the following parameters: $T_{SG} = 71.3$ K, $\tau_0^* = 9.1 \times 10^{-15}$ s, $zv = 8.7$, $T_0 = 68.8$ K, $\tau_0 = 9.3 \times 10^{-10}$ s, and $E_a/k_B = 87.4$ K, yielding $E_a/(k_B T_0) = 1.27$. This magnitude of $E_a/(k_B T_0) = 1.27 > 1$ along with $\Omega = 0.01$ places this system in the cluster SG category in disagreement with [41].

**2.2. Cluster Spin-glasses:**

**2.2.1. PrRhSn₃:** The analysis of Anand *et al* [19] for PrRhSn₃ using the temperature and frequency-dependent *ac* magnetic susceptibilities yielded $\Omega = 0.086$, $T_{SG} = 4.28$ K, $\tau_0^* = 2.04 \times 10^{-10}$ s, $zv = 10.9$, $T_0 = 4.01$ K, $\tau_0 = 4.7 \times 10^{-10}$ s and $E_a/k_B = 19.1$ K. Based on $\Omega = 0.086$, Anand *et al* characterized the system as a cluster SG. Fig. 3 shows the temperature dependence of the *ac* susceptibility extracted from their paper and in Fig. 4 we show the plots of our optimized analysis. Our analysis yielded $\Omega = 0.043$, optimum $T_{SG} = 4.20$ K with corresponding $\tau_0^* = 7.7 \times 10^{-10}$ s, and $zv = 4.7$, and optimum $T_0 = 3.8$ K, $\tau_0 = 3.9 \times 10^{-9}$ s and $E_a/k_B = 7.5$ K, the latter yielding $E_a/(k_B T_0) =$ 1.97 >1. The magnitudes of $zv$ and $E_a/k_BT_0$ are quite different from those reported in [19]. Based on the magnitude of $\Omega$ and $E_a/k_BT_0 = 1.97 >1$, we also classify this system as a cluster SG.

**2.2.2. $Cr_{0.5}Fe_{0.5}Ga$:** Bag *et al* investigated the SG behavior along with memory effects in $Cr_{0.5}Fe_{0.5}Ga$ [23] and reported $\Omega = 0.017$, $T_{SG} = 22.0$ K, $zv = 4.2$ and $\tau_0^* = 1.1 \times 10^{-10}$ s, $\tau_0 = 6.6 \times 10^{-9}$ s, $T_0 = 21.1$K and $E_a/k_B = 16.0$ K. Bag *et al.* classified $Cr_{0.5}Fe_{0.5}Ga$ as a cluster SG. Our analysis based on the *ac* susceptibility data of [23] and shown in Fig. S6 [34] yielded $\Omega = 0.045$, optimum $T_{SG} = 17.7$ K, $\tau_0^* = 1.1 \times 10^{-8}$ s, $zv = 3.5$, optimum $T_0 = 16.5$ K, $\tau_0 = 1.1 \times 10^{-7}$ s and $E_a/k_B = 17.2$ K (yielding $E_a/(k_BT_0) = 1.04 >1$). Considering $\Omega = 0.045$, and $E_a/(k_BT_0) = 1.04 >1$, we also classify this system as cluster SG. However, note the parameters given by Bag *et al* [23] yields $E_a/k_BT_0 = 0.76 < 1$.

**2.2.3. $Zn_3V_3O_8$:** Reporting on the bulk magnetic properties of vanadium-based geometrically frustrated system $Zn_3V_3O_8$, Chakrabarty *et al* noted that the SG state in $Zn_3V_3O_8$ originated from clusters of atoms rather than individual atoms and so identified the system as a cluster glass [42]. From the $\chi'$ vs $T$ data at different $f$ (Figure 4 of [42]), Chakrabarty *et al* obtained $\Omega = 0.028$, $T_{SG} = 3.75$ K, $\tau_0^* = 3 \times 10^{-3}$ s, $zv = 2.98$, $T_0 = 3.76$ K, $\tau_0 = 2.17 \times 10^{-4}$ s and $E_a/k_B = 0.36$ K, the latter yielding $E_a/(k_BT_0) = 0.096$. Our optimized analysis for the $\tau$ vs. $T_f$ data in $Zn_3V_3O_8$ (see Fig. S7 [34]) yielded $\Omega = 0.033$, optimum $T_{SG} = 3.6$ K, $\tau_0^* = 1.5 \times 10^{-9}$ s, $zv = 5.2$, optimum $T_0 = 3.3$ K, $\tau_0 = 1.6 \times 10^{-8}$ s and a substantially different value of $E_a/k_B = 6.3$ which in turn yields $E_a/(k_BT_0) = 1.91 >1$ quite different from the report in [42]. So our analysis provides a quantitative basis for classifying this system as a cluster SG.

**2.2.4. $ZnTiCoO_4$:** For this system, the analysis of the $\tau$ vs. $T_f$ data obtained from the peak in $\chi''$ was recently reported using our proposed method of analysis yielding $\Omega = 0.026$, and optimum $T_{SG} = 12.9$ K, $\tau_0^* = 1.0 \times 10^{-12}$ s and $zv = 11.75$, and $T_0 = 10.9$ K $\tau_0 = 1.6 \times 10^{-13}$ s, $E_a/k_BT_0 = 8.72$ [31]. Here we present results obtained from similar analysis of the $\tau$ vs. $T_f$ data obtained from the peaks in $\chi'$ and compare the values obtained in the two cases. This analysis, shown in Fig. S8 [34], yielded $\Omega = 0.033$, $T_{SG} = 13.2$ K, $\tau_0^* = 4.3 \times 10^{-11}$ s, $zv = 11.1$, optimum $T_0 = 10.9$ K, $\tau_0 = 2.1 \times 10^{-12}$ s and $E_a/(k_BT_0) = 9.44 > 1$. Although the optimum $T_0$ is identical in the two cases, magnitudes of the optimum $T_{SG}$ and all other parameters are slightly larger for the $\chi'$ data. This may be related to the fact that $T_f$ for $\chi'$ occurs at a slightly higher temperature than that for $\chi''$, and $T_f$ for $\chi''$ coincides with peak in $d(\chi'T)/dT$ [43]. However, both data sets yielded similar trends in their values and indicate towards the same conclusion that ZTCO is a cluster SG.

**2.2.5. Co₂RuO₄:** Ghosh *et al* reported magnetic investigations of the spinel $Co_2RuO_4$, which they classified as a cluster SG [21]. They analyzed the data of $\chi''$ vs T by employing the standard scaling laws and reported $\Omega = 0.01$, $T_{SG} = 14.97$ K, $\tau_0* = 1.16\times10^{-10}$ s, $zv = 5.2$, $T_0 = 14.3$ K and $\tau_0 = 1.1\times10^{-9}$ s. Our results from the new analysis are shown in Fig. S9 [34] yielding $\Omega = 0.018$, optimum $T_{SG} = 14.5$ K, with $\tau_0* = 2.6\times10^{-12}$ s, and $zv = 8.5$ and optimum $T_0 = 13.5$ K, $\tau_0 = 6.2\times10^{-10}$ s and $E_a/(k_BT_0) = 2.55 > 1$. Using the criteria of $E_a/(k_BT_0) > 1$, $Co_2RuO_4$ is a cluster SG, in agreement with the conclusion in [21] although the magnitudes of the optimum parameters determined here are different from those reported in [21].

**2.2.6. Pr₄Ni₃O₈:** In $Pr_4Ni_3O_8$, Huangfu *et al* [44] reported SG behavior along with magnetic memory effects between 2 and 300 K. They analyzed the $\tau$ vs. $T_f$ data from the $\chi'$ vs. T plots (figure 3 (a) from [44]) and reported $T_{SG}$ = 68.3 K, $zv$ = 3.8 and $\tau_0^*$ ~ $10^{-6}$ s, $T_0$ = 60.1 K, $\tau_0$ ~ $10^{-6}$ s, and $E_a/k_B$ ~ 130 K, leading them to categorize $Pr_4Ni_3O_8$ as a SG due to short-range magnetic ordering. Our analysis establishes $Pr_4Ni_3O_8$ as a cluster SG based on the following optimized parameters: $\Omega$ = 0.066, $T_{SG}$ = 75.0 K, $\tau_0^*$ = 2.5×$10^{-6}$ s, $zv$ = 3.4, optimum $T_0$ = 67.0 K, $\tau_0$ = 4.4×$10^{-6}$ s and $E_a/(k_BT_0)$ = 1.67. (see Fig. S10 [34]). Note the different magnitudes of the optimized parameters vis-a-vis those reported in [44].

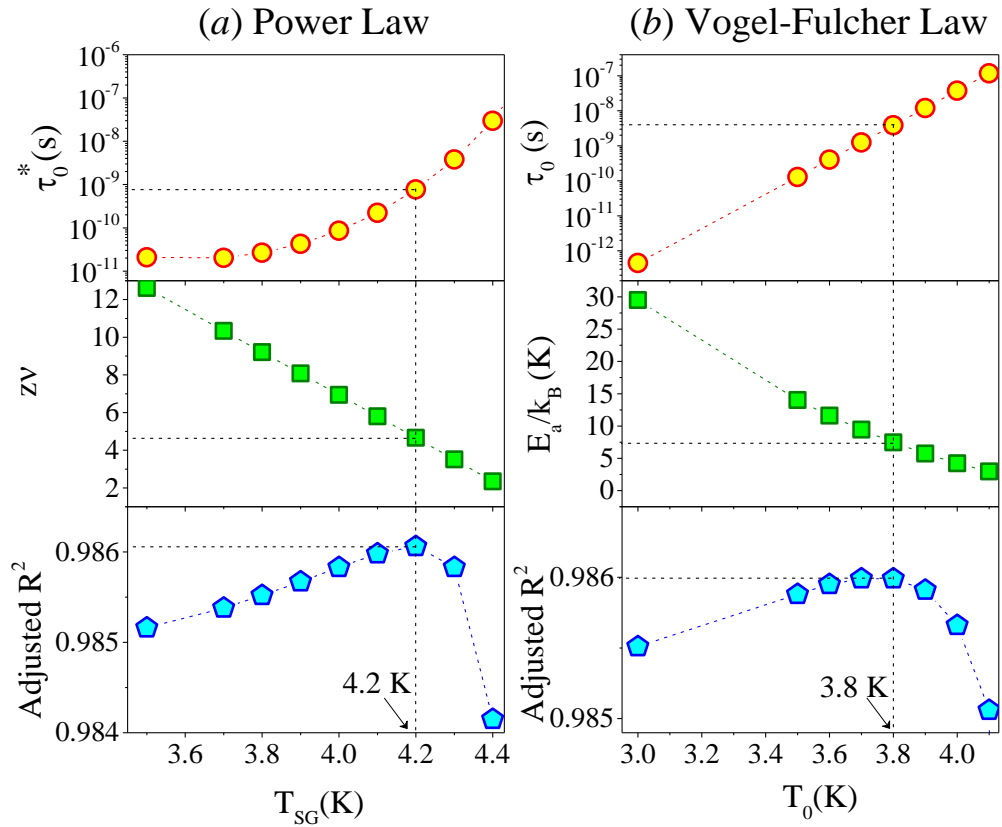

**Figure 4:** Same as Fig. 2 except the plots and analysis are for PrRhSn₃ using the information from Fig. 3.

### 2.2.7. $\beta$-MoCl$_4$:

McGuire *et al* reported experimental studies on the layered $\beta$-MoCl$_4$ and commented on its SG nature [45]. However, they did not characterize the system in detail using the conventional dynamical scaling laws. By extracting their data for the frequency dependence of $\chi'$ vs. T given in figure 4 (c) in [45], we performed the analysis shown in Fig. S11 [34] yielding $\Omega$ = 0.021, optimum $T_{SG}$ = 4.8 K with $\tau_0^*$ = 7.3×$10^{-15}$ s, $zv$ = 10.4, optimum $T_0$ = 4.3 K, $\tau_0$ = 1.0×$10^{-13}$ s and $E_a/(k_BT_0)$ = 4.84 > 1. Using $\Omega$ = 0.021 and $E_a/(k_BT_0)$ > 1 from our analysis, we suggest $\beta$-MoCl$_4$ to be a cluster SG.

### 2.2.8. $Zn_{0.5}Ni_{0.5}Fe_2O_4$:

Botez *et al* [46] reported magnetic properties of $Zn_{0.5}Ni_{0.5}Fe_2O_4$ nanoparticles with average size ~ 9 nm and observed a bifurcation of the zero-field-cooled and field-cooled dc magnetic susceptibilities below ~ 190 K, as observed in magnetic nanoparticles [47]. But their analysis of the *ac* susceptibilities near 190 K yielded $\Omega$ = 0.032 which is typical of cluster SG systems, and the use of VFL and PL analysis yielded $T_{SG}$ = 190 K, $T_0$ = 180 K, $\tau_0$ = $10^{-11}$ s and $zv$ = 10.0. These results combined with $H^{2/3}$ dependence of $\Delta T_{SG}$ following the Almieda-Thouless line led them to conclude that $T_{SG}$ is due to SG-like collective freezing of super-spins of the nanoparticles [46]. Our optimized analysis of their $\chi'$ vs. T data (shown in Fig. S12 [34]) yields $\Omega$ = 0.03, optimum $T_{SG}$ = 197 K, $\tau_0^*$ = 6.1×$10^{-12}$ s, $zv$ = 7.6, optimum $T_0$ = 177 K, $\tau_0$ = 4.9×$10^{-11}$ s, $E_a/k_B$ = 612 K, and $E_a/k_BT_0$ = 3.46 >1. So, this system is classified as a cluster SG showing that strongly interacting magnetic nanoparticles can also have characteristics of cluster SG.

### 2.2.9. LiFeSnO$_4$-LT:

For the low-temperature (LT) phase of LiFeSnO$_4$ with the distorted Kagome lattice, Banerjee *et al* [40] reported $\tau_0^*$ = 9×$10^{-10}$ s, $zv$ = 4.0, $T_0$ =15.6 K, $\Omega$ = 0.018 and $E_a/k_B$ = 24.7 K yielding $E_a/(k_BT_0)$ = 1.6. They classified the LT phase as cluster SG based on the larger value of $\tau_0^*$ = 9×$10^{-10}$ s compared to $\tau_0^*$ = 3×$10^{-13}$ s for the HT phase even though magnitudes of $\Omega = 0.018$ are nearly identical for the HT and LT phases. Our analysis of the data for the LT phase extracted from [40] (shown in Fig. S13 [34]) yielded optimum $T_{SG} = 15.0$ K, $\tau_0* = 9.6 \times 10^{-20}$ s, $zv$ = 19.2, $T_0 = 12.7$ K, $\tau_0 = 8.0 \times 10^{-21}$ s and $E_a/k_B = 177.4$ K. Since $E_a/(k_B T_0) = 14.0 > 1$, the LT phase is classified as cluster SG.

**2.2.10. ScEr**: Here we report results from our optimized analysis of the $T_f$ vs. $f$ data taken from the *ac* susceptibility studies of Wendler *et al* [48]. For ScEr$_{10\%}$, this analysis shown in S14 [34] yielded $\Omega = 0.045$, optimum $T_{SG} = 2.5$ K, $\tau_0* = 4.8 \times 10^{-10}$ s, $zv$ =13.5, optimum $T_0 = 1.6$ K, $\tau_0 = 6.4 \times 10^{-16}$ s, $E_a/k_B = 49.2$ K leading to $E_a/k_B T_0 = 30.8 > 1$. Based on the magnitudes of $E_a/k_B T_0$ and $\Omega$, this system is classified as cluster SG. For comparison, Wendler *et al* [48] reported $T_{SG} = 2.8$ K, $zv$ =9.8, $\tau_0* = 6.7 \times 10^{-9}$ s, $T_0 = 2.16$ K, and $\tau_0 \sim 10^{-10}$ s. Interestingly $\tau_0 < \tau_0*$ in this case.

**2.2.11. ScDy**: The data for this system (ScDy$_{3.5\%}$) is also taken from [48] and our optimized analysis shown in Fig. S15 [34] yielded $T_{SG} = 2.9$ K, $\tau_0* = 3 \times 10^{-7}$ s, $zv = 10.4$, $T_0 = 1.9$ K, $\tau_0 = 6.9 \times 10^{-12}$ s and $E_a/k_B = 43.5$ K leading to $E_a/k_B T_0 = 22.9 > 1$. So, this system is also a cluster SG based on $E_a/k_B T_0 > 1$ and $\Omega = 0.076$. In this case also, $\tau_0 < \tau_0*$.

**3. Interpretation**

To explain the result of $T_{SG} > T_0$ evident from our optimized parameters listed in Table 1, Eqs. (2) and (3) are rewritten as:

$$Ln(\tau/\tau_o) = E_a/k_B(T_f - T_0) \qquad ---(4)$$

$$Ln(\tau/\tau_o*) = - zv \, Ln \, [(T_f - T_{SG})/T_{SG}]. \qquad -----(5)$$

Following Souletie and Tholence [37], we use $Ln(\tau/\tau_o) \sim 20$ and $Ln(\tau/\tau_o*) \sim 25$. Eqs. (4) and (5) can then be solved yielding

$$(T_{SG} - T_0) = (E_a/20k_B) - T_{SG} \, \exp(-25/zv). \qquad ------ (6).$$

From Eq. (6), $(T_{SG} - T_0)$ can be calculated using optimized values of $E_a/k_B$, $T_{SG}$ and $zv$ listed in Table 1 for each case and compared with the optimized fitted value of $(T_{SG} - T_0)$. This comparison of the "experimental" and "calculated" values of $(T_{SG} - T_0)$ for all the systems discussed here are also given in Table I, showing that the calculated $(T_{SG} - T_0)$ is indeed positive, meaning $T_{SG} > T_0$ although the agreement between the "experimental" and "calculated" values of $(T_{SG} - T_0)$ is better in some cases than in other cases. This difference is likely related to some difference in the approximate values of $Ln(\tau/\tau_o) \sim 20$ and $Ln(\tau/\tau_o*) \sim 25$ used here for all cases in the calculations.

This explanation of the result $T_{SG} > T_0$ provides additional confidence in the correctness of the optimized analysis presented in this paper.

The optimized analysis in seventeen SG systems presented here shows that canonical and cluster SG systems have different magnitudes of the ratio $E_a/(k_B T_0)$ and the Mydosh parameter $\Omega$ in that $E_a/(k_B T_0) < 1$ along with $\Omega < 0.01$ for canonical SG, and $E_a/(k_B T_0) > 1$ along with $\Omega > 0.01$ for cluster SG. In contrast, the associated magnitudes of $\tau_0{}^*$, $\tau_0$, and $zv$ are found to be less reliable in distinguishing between different SG system since no systematic pattern from Table I is evident in their variations between canonical and cluster SG. All SG systems consist of spin clusters [8,9], the difference between canonical and cluster SG is the size of the clusters and energy barrier between the ground state valleys of various clusters. The condition $E_a/(k_B T_0) < 1$ implies that in canonical SG, the energy barrier $E_a/k_B$ is smaller than the inter cluster coupling $T_0$ whereas for $E_a/(k_B T_0) > 1$ valid for cluster SG, $E_a/k_B > T_0$.

## 4. Concluding Remarks:

In this paper, we have presented results using a new optimized procedure for fitting the data of $\tau$ vs. $T_f$ to the VFL (Eq. 2) and the PL (Eq. 3) for seventeen SG systems for which we could extract such data from published sources. This analysis showed how the parameters $\tau_0$ and $E_a/k_B$ for the VFL and $\tau_0{}^*$ and $zv$ for the PL strongly depend on the choice of $T_0$ and $T_{SG}$ respectively. Hence, to eliminate this uncertainty, we employed the maximum AR$^2$ method to determine the optimum values of $T_0$ and $T_{SG}$, and it is suggested that the magnitudes $\tau_0$, $E_a/k_B$, $\tau_0{}^*$ and $zv$ associated with these optimum values of $T_{SG}$ and $T_0$ should be used in the discussion, analysis, and classification of SG systems. Results from this analysis show that magnitudes of $\tau_0{}^*$, $\tau_0$ and $zv$ fail to provide a clear distinction between canonical and cluster SG. However, new results emerge showing $E_a/(k_B T_0) < 1$ in canonical SG systems, and $E_a/(k_B T_0) > 1$ for cluster SG systems. Also, the optimized $T_{SG} >$ optimized $T_0$ in all cases. More rigorous theoretical interpretation of the boundary near $E_a/(k_B T_0) \sim 1$ is desired along with testing of this criterion for distinguishing between canonical and cluster SG systems as appropriate data in such systems become available in the future.

## Supplementary Material:

The Supplementary material contains graphs representing the optimized procedure for the analysis of $\tau$ vs. $T_f$ data in fifteen SG systems, excluding that for AuMn$_{2.98\%}$ and PrRhSn$_3$, analysed in this work. The Supplementary Material has been assigned the reference number [34] for use in the main text.

**Please Note:**

The following article has been submitted to AIP Advances.

**Acknowledgements:**

M.R.C. acknowledges the Ministry of Education, Government of India for providing financial support for her research work through the Prime Minister's Research Fellowship May 2021 scheme. M.R.C. and S.T. acknowledge the FIST program of the Department of Science and Technology, India for partial support of this work (Grants No. SR/FST/PSII-020/2009 and No. SR/FST/PSII-037/2016). M.R.C. and S.T. acknowledge the financial support from UGC-DAE CSR through a Collaborative Research Scheme (CRS) Project Number CRS/2021-22/01/383. S.T. acknowledges the DST SERB Core Research Grant File No. CRG/2022/006155 for the partial support of this work. M.R.C. and S.T. acknowledge the Central Instrument Facility of the Indian Institute of Technology, Guwahati for support. M.R.C. expresses her sincere thanks to Dr. Sayandeep Ghosh for useful discussions.

**Table 1:** Optimum magnitudes of the various parameters for seventeen systems determined from fits to Eqs. (1), (2) and (3).

| Compound | $\Omega$ | Power Law Analysis | | | Vogel-Fulcher Law Analysis | | | | Exp. $(T_{SG}-T_0)$ (K) | Calc. $(T_{SG}-T_0)$ (K) |
|---|---|---|---|---|---|---|---|---|---|---|
| | | $T_{SG}$ (K) | $\tau_0^*$ (s) | $z\upsilon$ | $T_0$ (K) | $\tau_0$ (s) | $E_a/k_B$ (K) | $E_a/k_BT_0$ | | |
| $AuMn_{2.98\%}$ | 0.006 | 10.0 | $3.6\times10^{-16}$ | 7.6 | 9.7 | $1.7\times10^{-11}$ | 9.5 | 0.98 | 0.3 | 0.10 |
| $CuMn_{4.6\%}$ | 0.008 | 27.3 | $4.8\times10^{-15}$ | 6.8 | 26.7 | $8.8\times10^{-10}$ | 16.7 | 0.62 | 0.6 | 0.14 |
| $Na_{0.7}MnO_2$ | 0.005 | 38.6 | $1.5\times10^{-14}$ | 6.0 | 38.1 | $1.3\times10^{-8}$ | 12.3 | 0.32 | 0.5 | 0.02 |
| $CaSrFeRuO_6$ | 0.009 | 63.6 | $6.2\times10^{-9}$ | 2.0 | 63.2 | $2.1\times10^{-6}$ | 3.3 | 0.05 | 0.4 | 0.17 |
| $LiFeSnO_4$-HT | 0.009 | 21.3 | $3.6\times10^{-9}$ | 2.8 | 20.8 | $3.6\times10^{-7}$ | 5.7 | 0.27 | 0.5 | 0.28 |
| $IrMnGa$ | 0.01 | 71.3 | $9.1\times10^{-15}$ | 8.7 | 68.8 | $9.3\times10^{-10}$ | 87.4 | 1.27 | 2.5 | 0.39 |
| $PrRhSn_3$ | 0.043 | 4.2 | $7.7\times10^{-10}$ | 4.7 | 3.8 | $3.9\times10^{-9}$ | 7.5 | 1.97 | 0.4 | 0.35 |
| **$Cr_{0.5}Fe_{0.5}Ga$** | 0.045 | 17.7 | $1.1 \times 10^{-8}$ | 3.5 | 16.5 | $1.1 \times 10^{-7}$ | 17.2 | 1.04 | 1.2 | 0.85 |
| **$Zn_3V_3O_8$** | 0.033 | 3.6 | $1.5 \times 10^{-9}$ | 5.2 | 3.3 | $1.6 \times 10^{-8}$ | 6.3 | 1.91 | 0.3 | 0.29 |
| **$ZnTiCoO_4$ ($\chi'$)** | 0.033 | 13.2 | $4.3 \times 10^{-11}$ | 11.1 | 10.9 | $2.1 \times 10^{-12}$ | 103.0 | 9.45 | 2.3 | 3.76 |
| **$ZnTiCoO_4$ ($\chi''$)** | 0.026 | 12.9 | $1.0 \times 10^{-12}$ | 11.8 | 10.9 | $1.6 \times 10^{-13}$ | 95.1 | 8.72 | 2.0 | 3.20 |
| **$Co_2RuO_4$** | 0.018 | 14.5 | $2.6 \times 10^{-12}$ | 8.5 | 13.5 | $6.2 \times 10^{-10}$ | 34.5 | 2.55 | 1.0 | 0.95 |
| **$Pr_4Ni_3O_8$** | 0.066 | 75.0 | $2.5 \times 10^{-6}$ | 3.4 | 67.0 | $4.4 \times 10^{-6}$ | 112.0 | 1.67 | 8.0 | 5.55 |
| **$\beta\text{-}MoCl_4$** | 0.021 | 4.8 | $7.3 \times 10^{-15}$ | 10.4 | 4.3 | $1.0 \times 10^{-13}$ | 20.8 | 4.84 | 0.5 | 0.60 |
| **$Zn_{0.5}Ni_{0.5}Fe_2O_4$** | 0.030 | 197.0 | $6.1 \times 10^{-12}$ | 7.6 | 177.0 | $4.9 \times 10^{-11}$ | 612.0 | 3.46 | 20.0 | 27.3 |
| **$LiFeSnO_4\text{-}LT$** | 0.017 | 15.0 | $9.6 \times 10^{-20}$ | 19.2 | 12.7 | $8.0 \times 10^{-21}$ | 177.4 | 14.0 | 2.3 | 4.78 |
| **$ScEr_{10\%}$** | 0.045 | 2.5 | $4.8 \times 10^{-10}$ | 13.5 | 1.6 | $6.4 \times 10^{-16}$ | 49.2 | 30.8 | 0.9 | 2.07 |
| **$ScDy_{3.5\%}$** | 0.076 | 2.9 | $3.0 \times 10^{-7}$ | 10.4 | 1.9 | $6.9 \times 10^{-12}$ | 43.5 | 22.9 | 1.0 | 1.91 |

**Figure S1:** Like the plots shown in Fig. 2 for AuMn$_{2.98\%}$ in the paper, shown here are corresponding plots for CuMn$_{4.6\%}$ on the variation of the fitting parameters and adjusted R$^2$ with changes in $T_{SG}$ for the Power Law (Eq. 3) and $T_0$ for the Vogel-Fulcher Law (Eq. 2). The vertical dotted lines mark the maximum value of adjusted R$^2$ yielding the corresponding optimum magnitudes of the various parameters listed in Table 1.

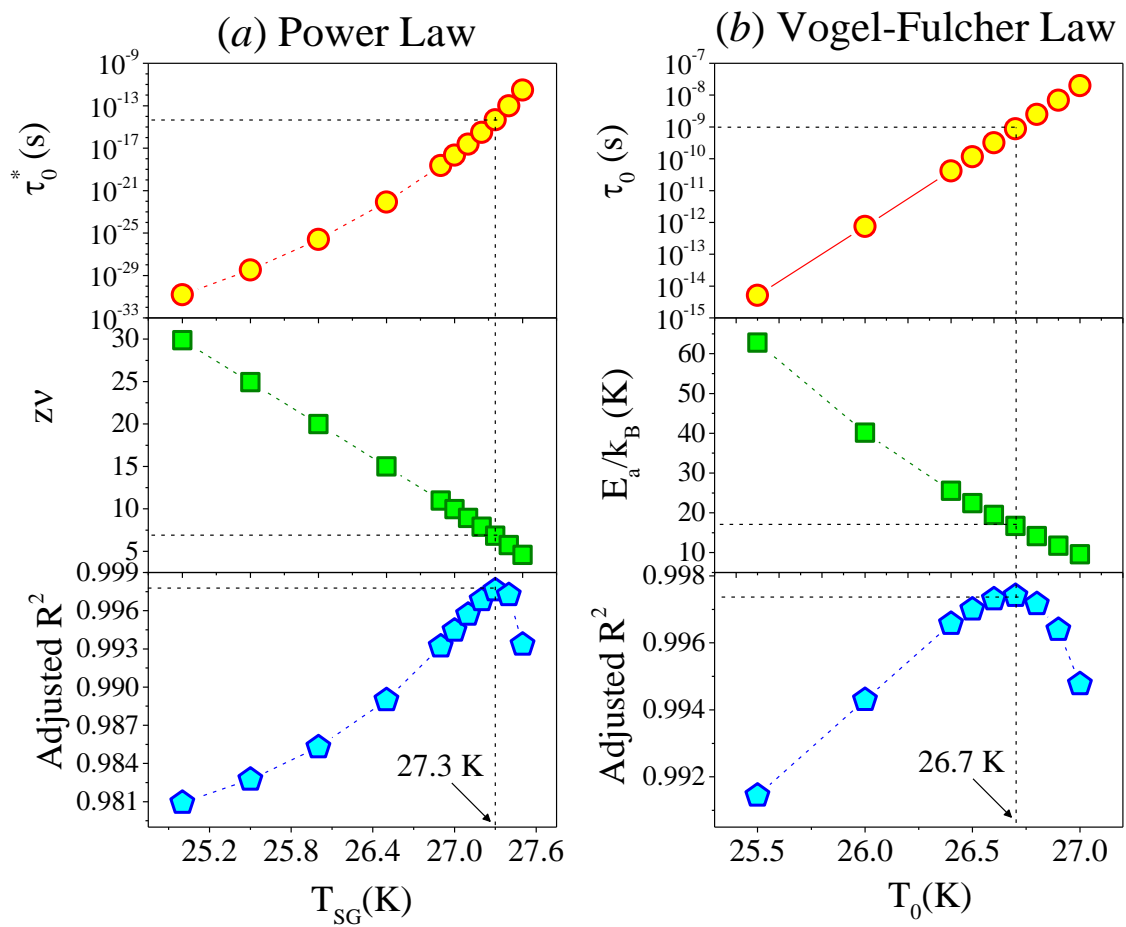

**Figure S2:** Same as for Fig. S1 except the plots and analysis are for Na$_{0.7}$MnO$_2$.

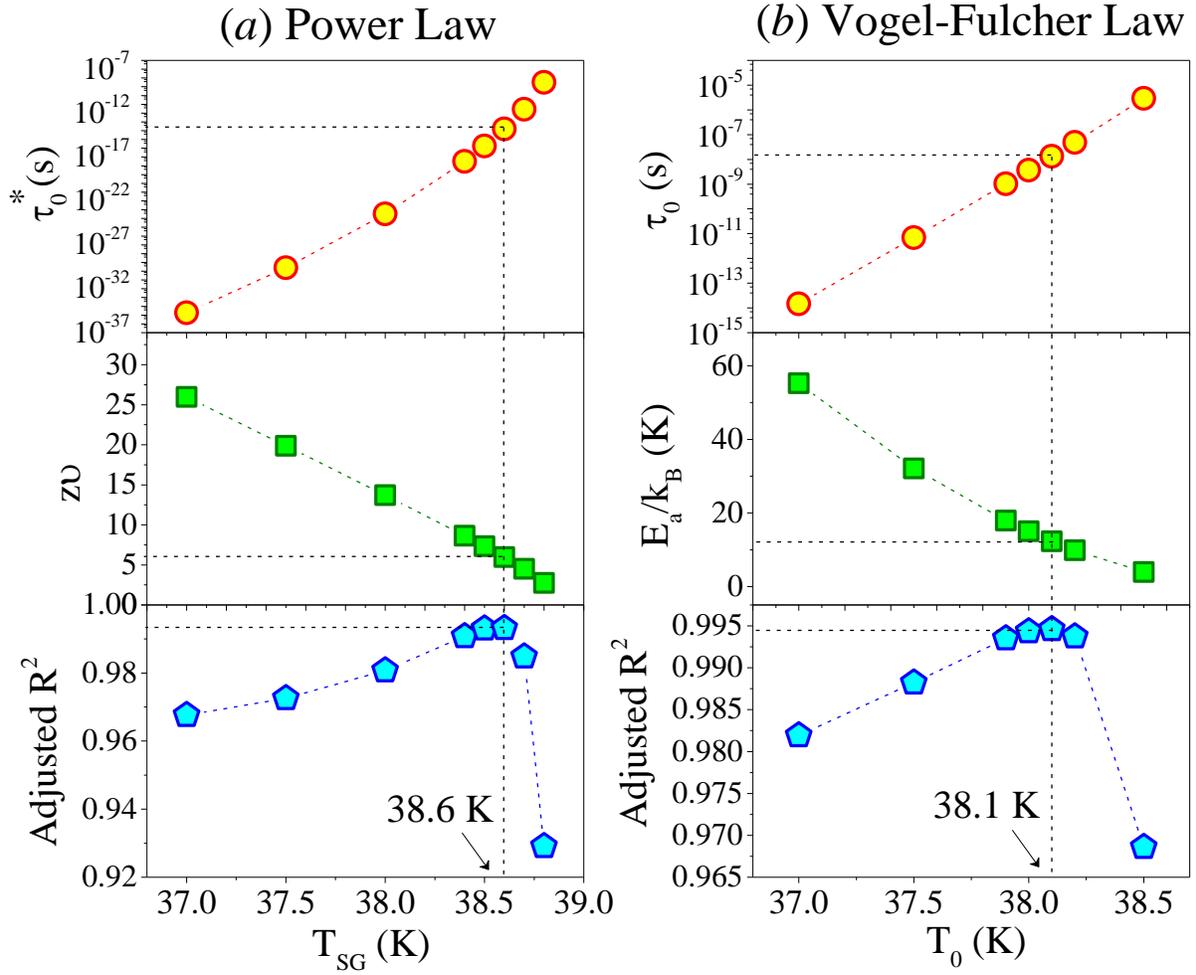

## (*a*) Power Law      (*b*) Vogel-Fulcher Law

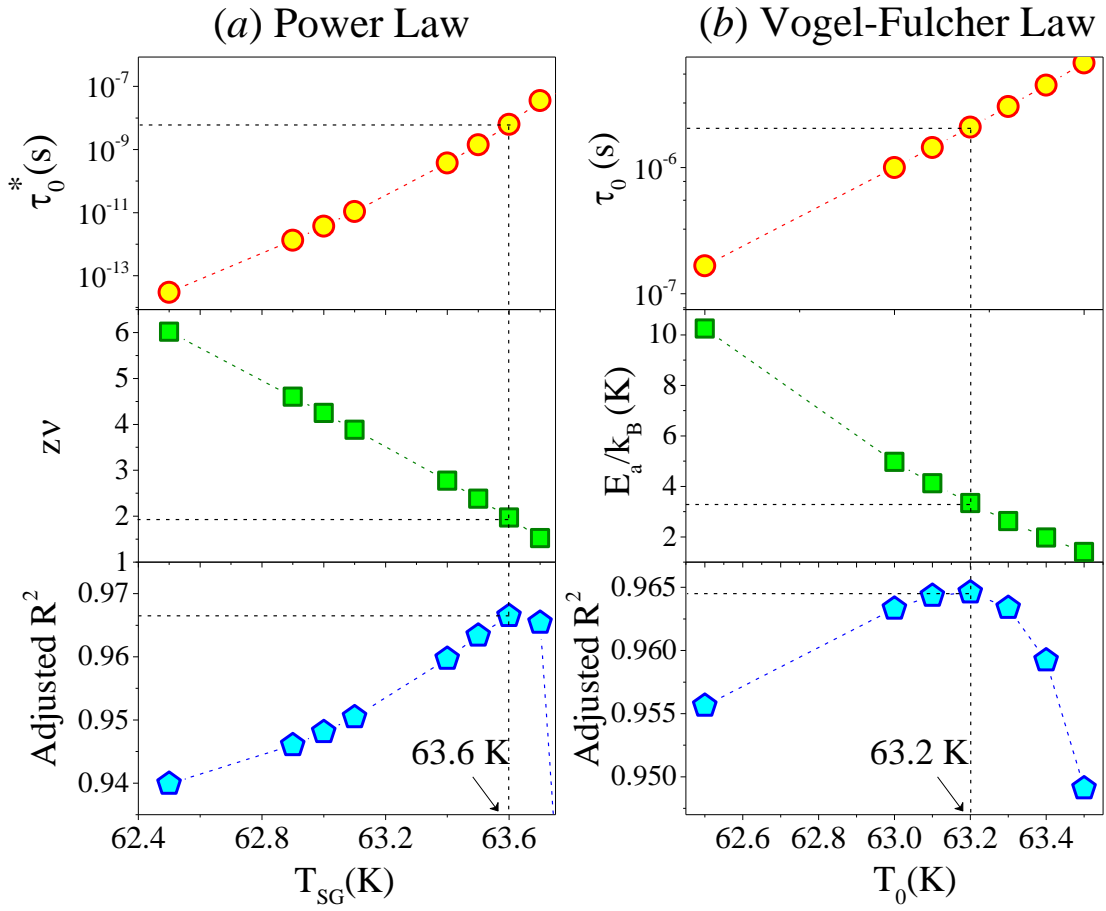

**Figure S4:** Same as for Fig. S1 except the plots and analysis are for LiFeSnO$_4$-HT.

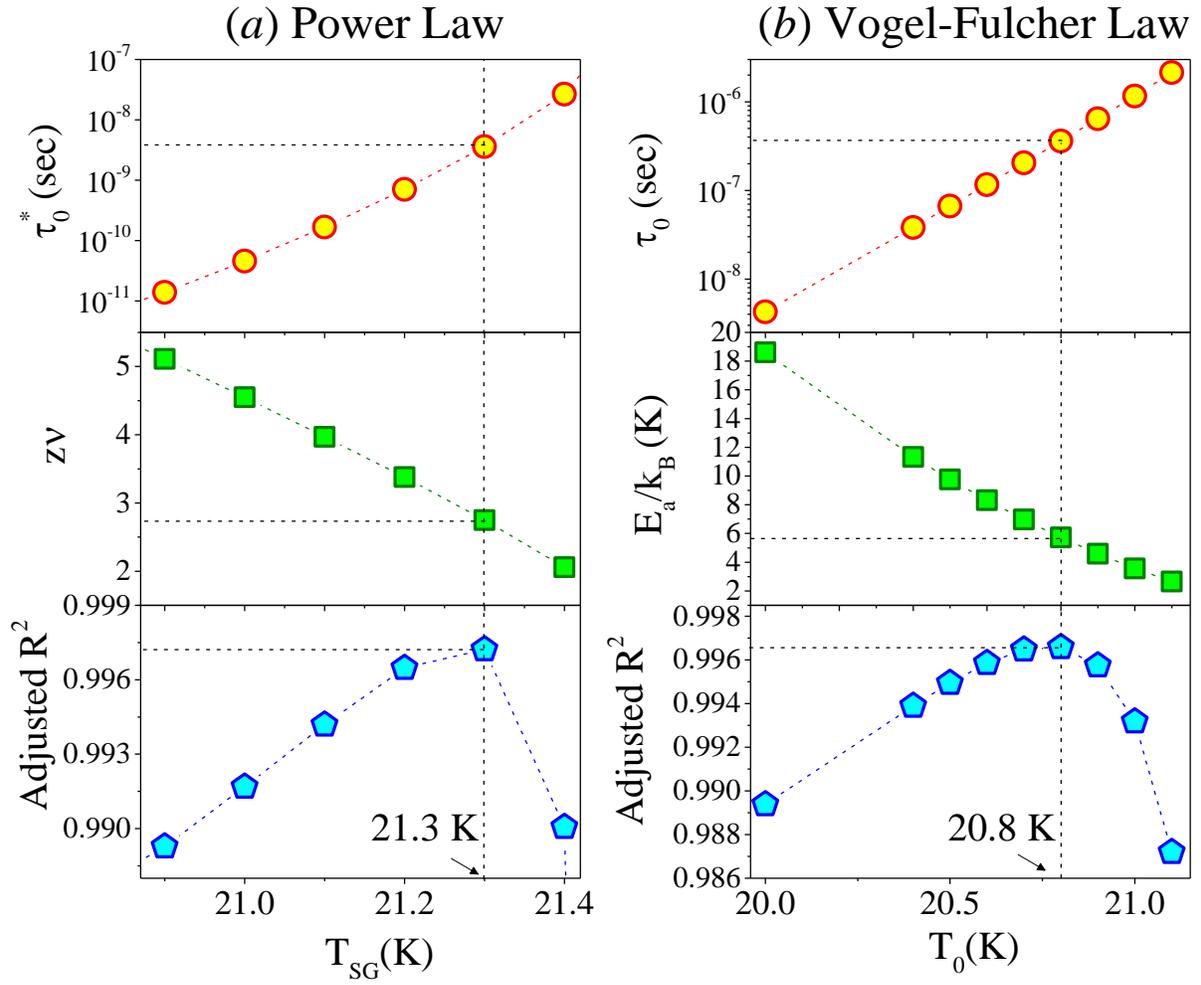

(*a*) Power Law      (*b*) Vogel-Fulcher Law

**Figure S5:** Same as for Fig. S1 except the plots and analysis are for IrMnGa.

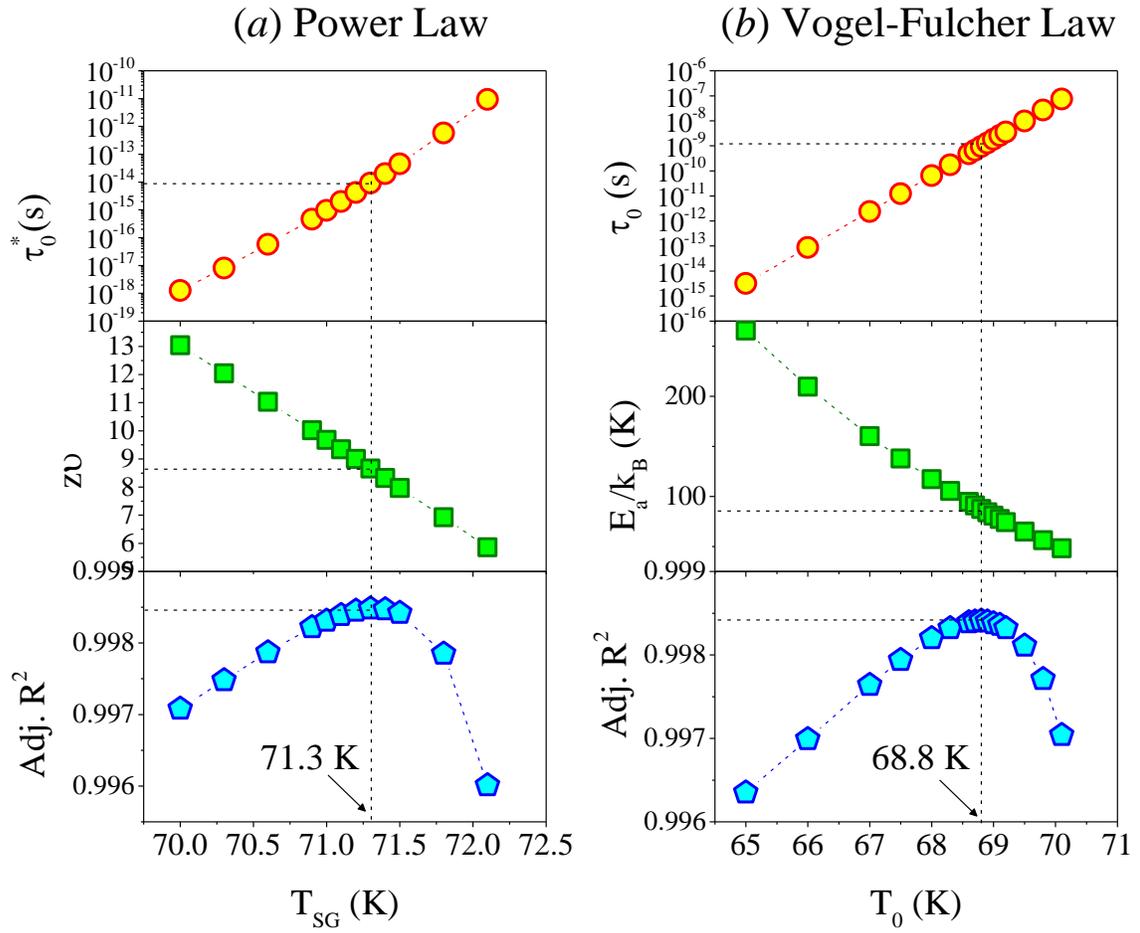

(*a*) Power Law   (*b*) Vogel-Fulcher Law

**Figure S6:** Same as for Fig. S1 except the plots and analysis are for $Cr_{0.5}Fe_{0.5}Ga$.

(*a*) Power Law　　　(*b*) Vogel-Fulcher Law

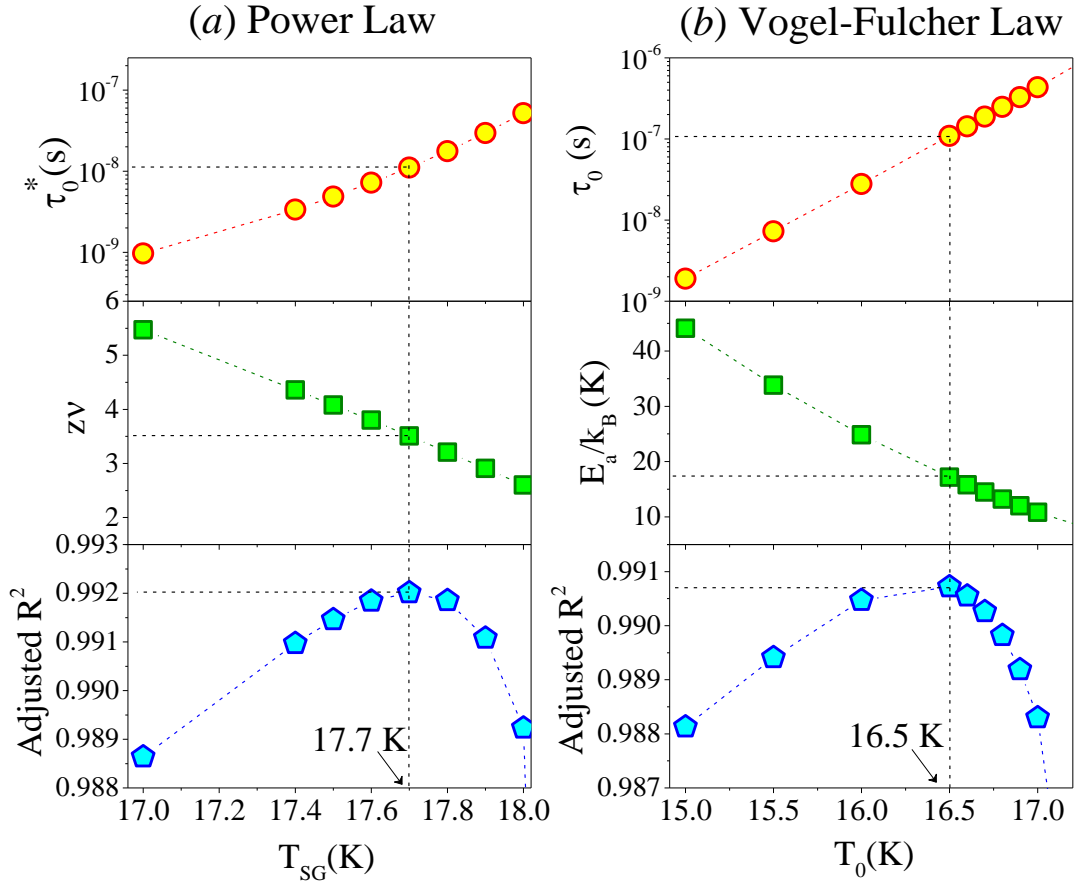

**Figure S7:** Same as for Fig. S1 except the plots and analysis are for $Zn_3V_3O_8$.

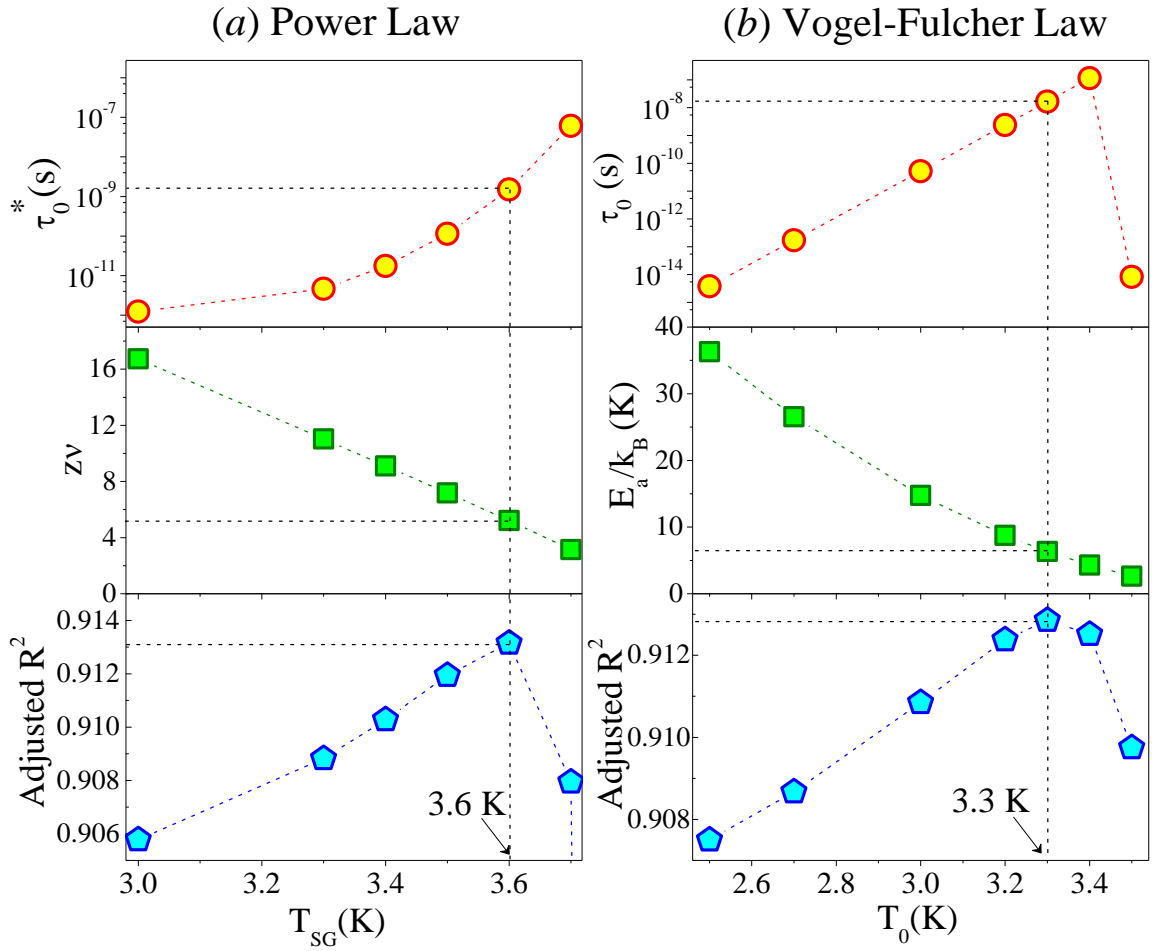

**Figure S8:** Same as for Fig. S1 except the plots and analysis are for ZnTiCoO₄.

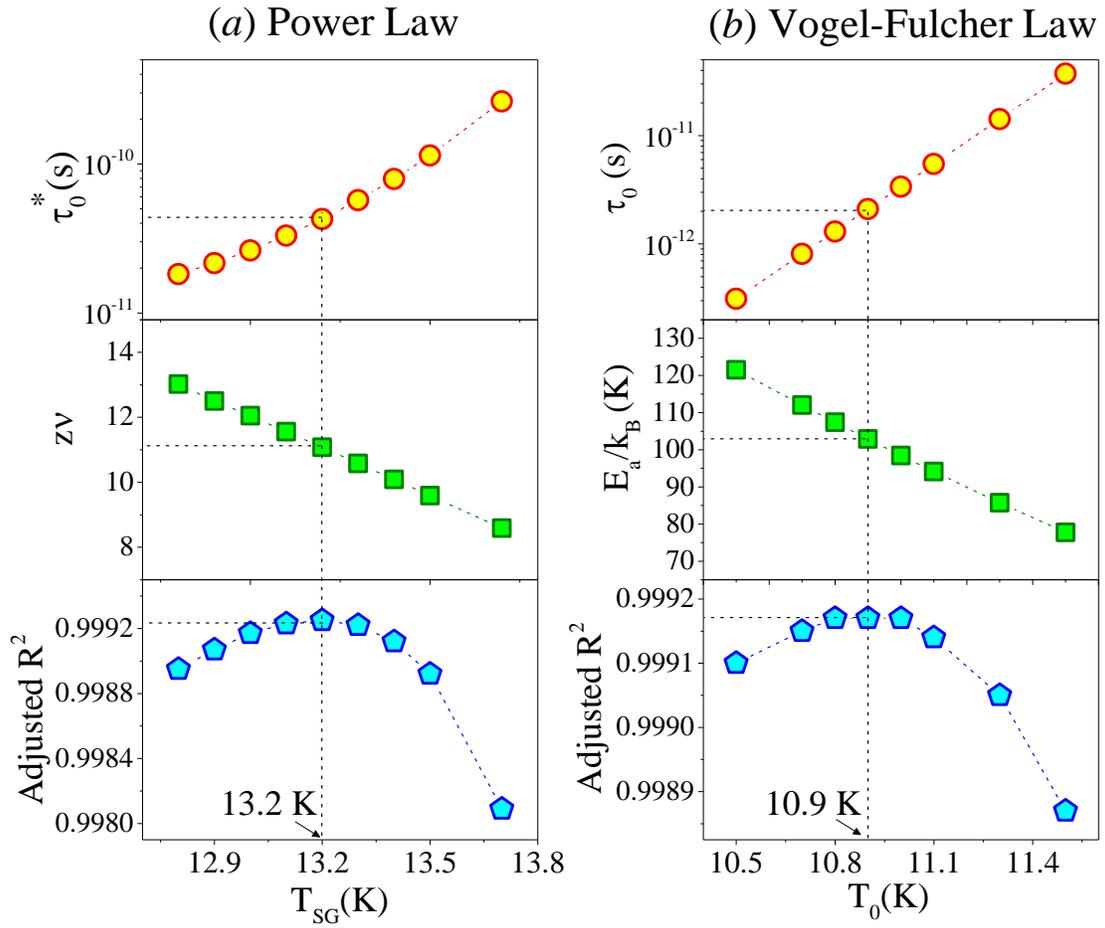

**Figure S9:** Same as for Fig. S1 except the plots and analysis are for $Co_2RuO_4$.

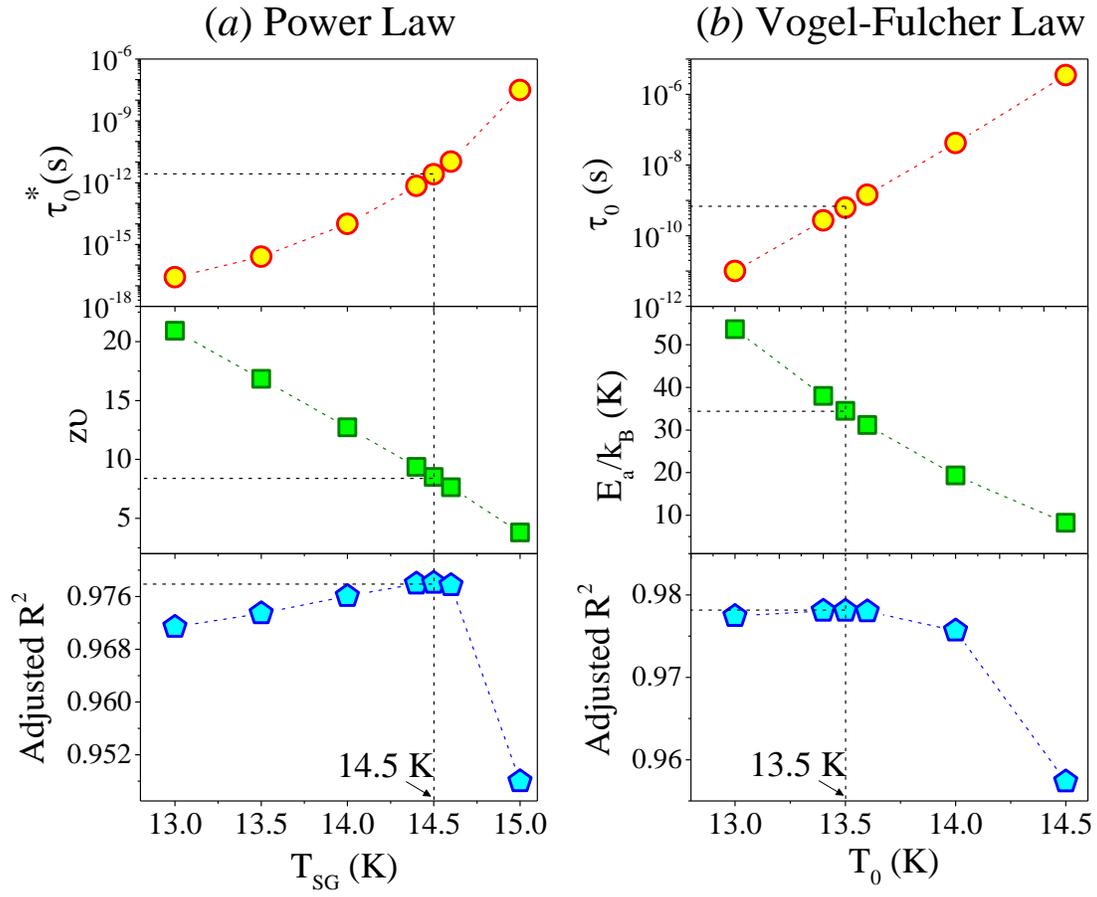

(*a*) Power Law        (*b*) Vogel-Fulcher Law

**Figure S10:** Same as for Fig. S1 except the plots and analysis are for Pr₄Ni₃O₈.

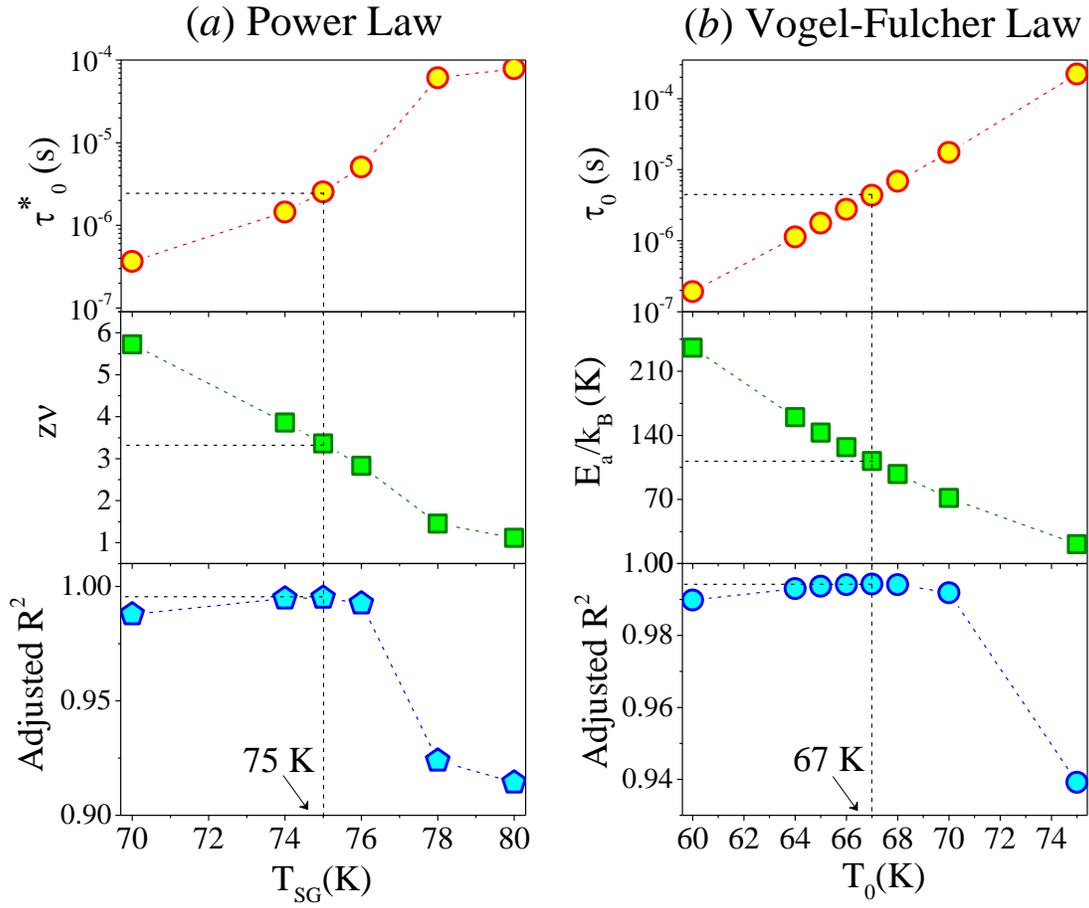

**Figure S11:** Same as for Fig. S1 except the plots and analysis are for $\beta$-MoCl$_4$.

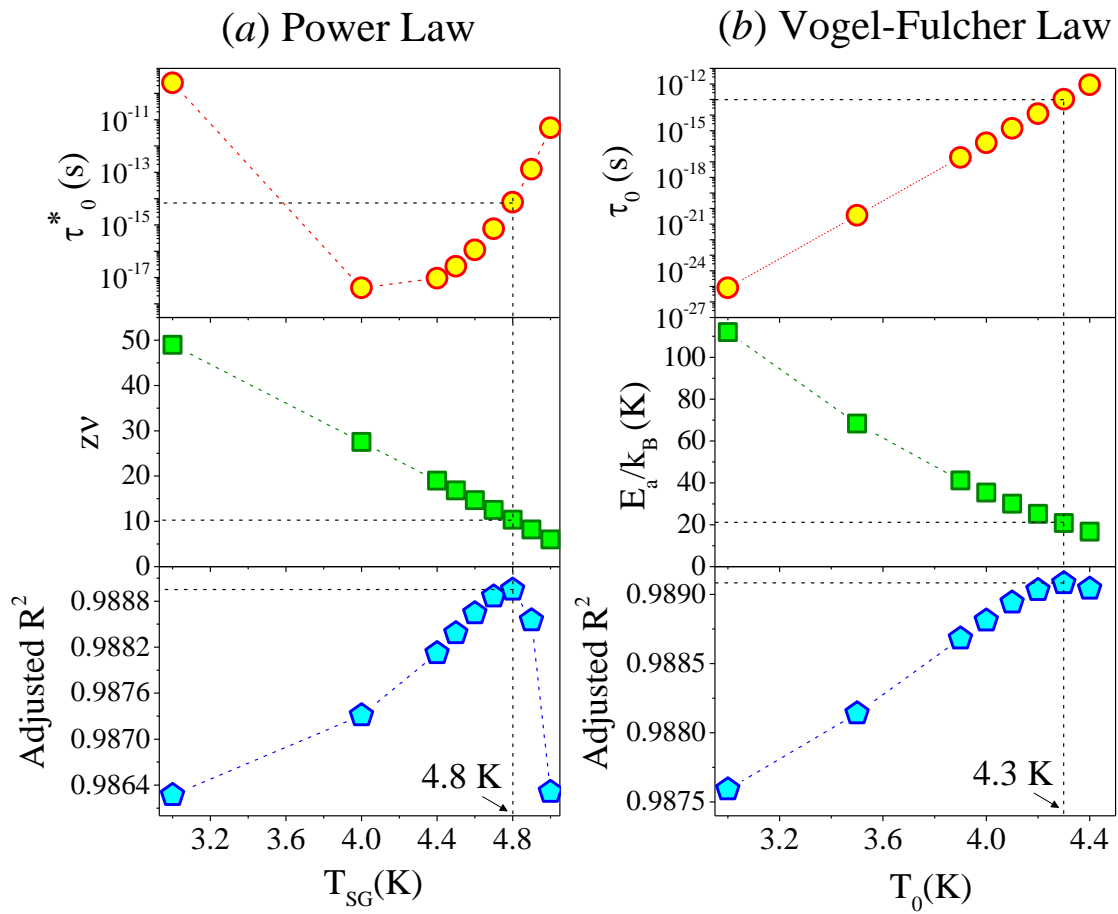

**Figure S12:** Same as for Fig. S1 except the plots and analysis are for $Zn_{0.5}Ni_{0.5}Fe_2O_4$.

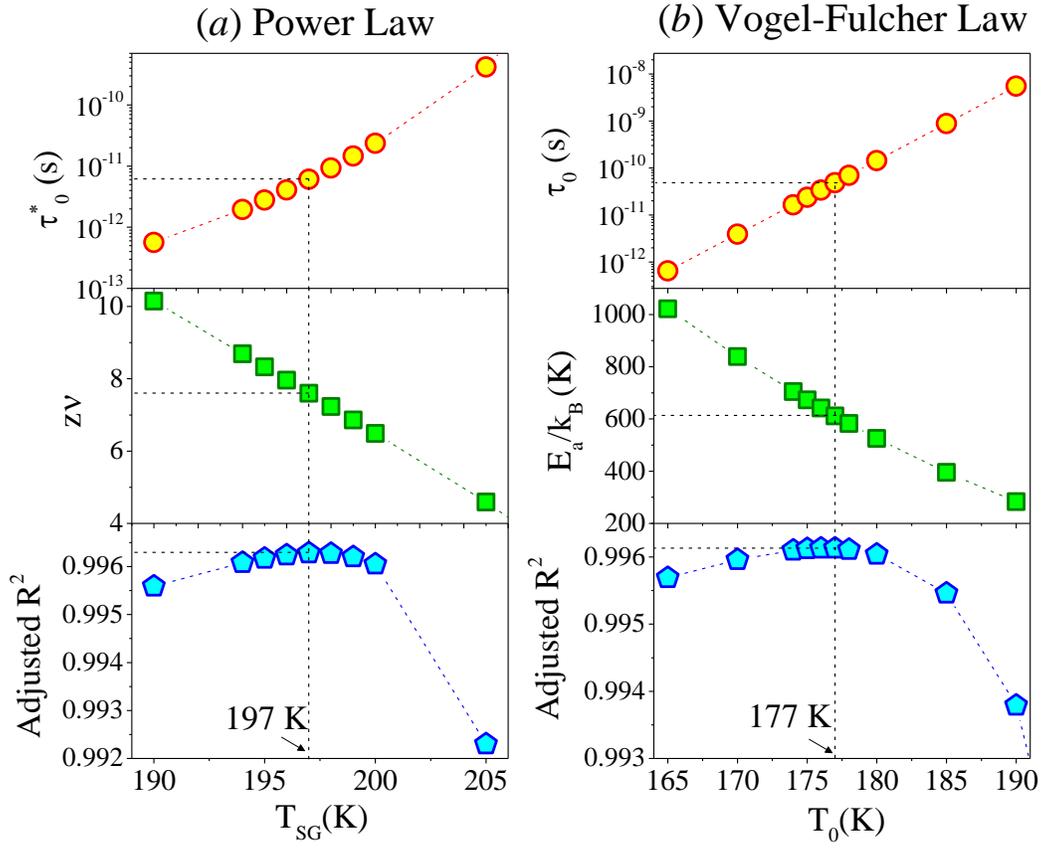

**Figure S13:** Same as for Fig. S1 except the plots and analysis are for LiFeSnO₄-LT.

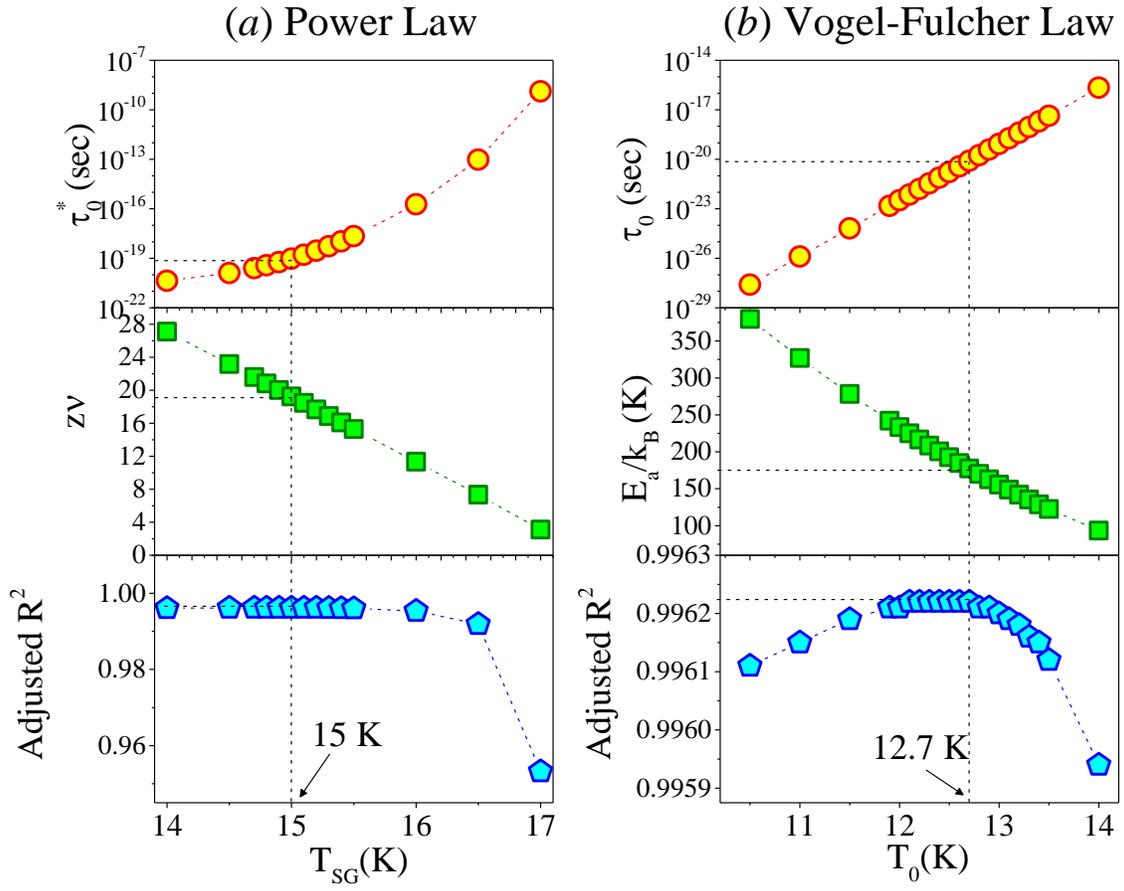

**Figure S14:** Same as for Fig. S1 except the plots and analysis are for ScEr_{10%}.

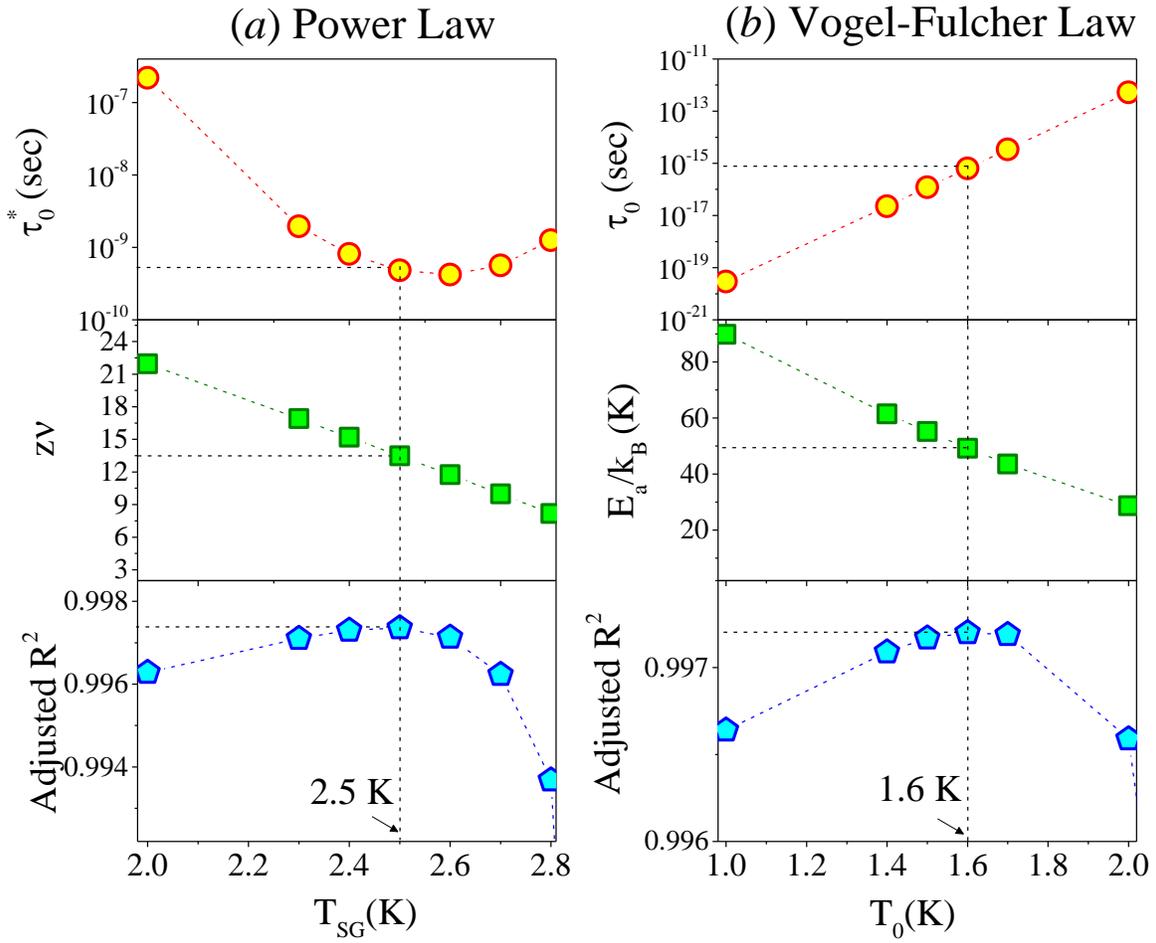

(*a*) Power Law          (*b*) Vogel-Fulcher Law

**Figure S15:** Same as for Fig. S1 except the plots and analysis are for ScDy₃.₅%.

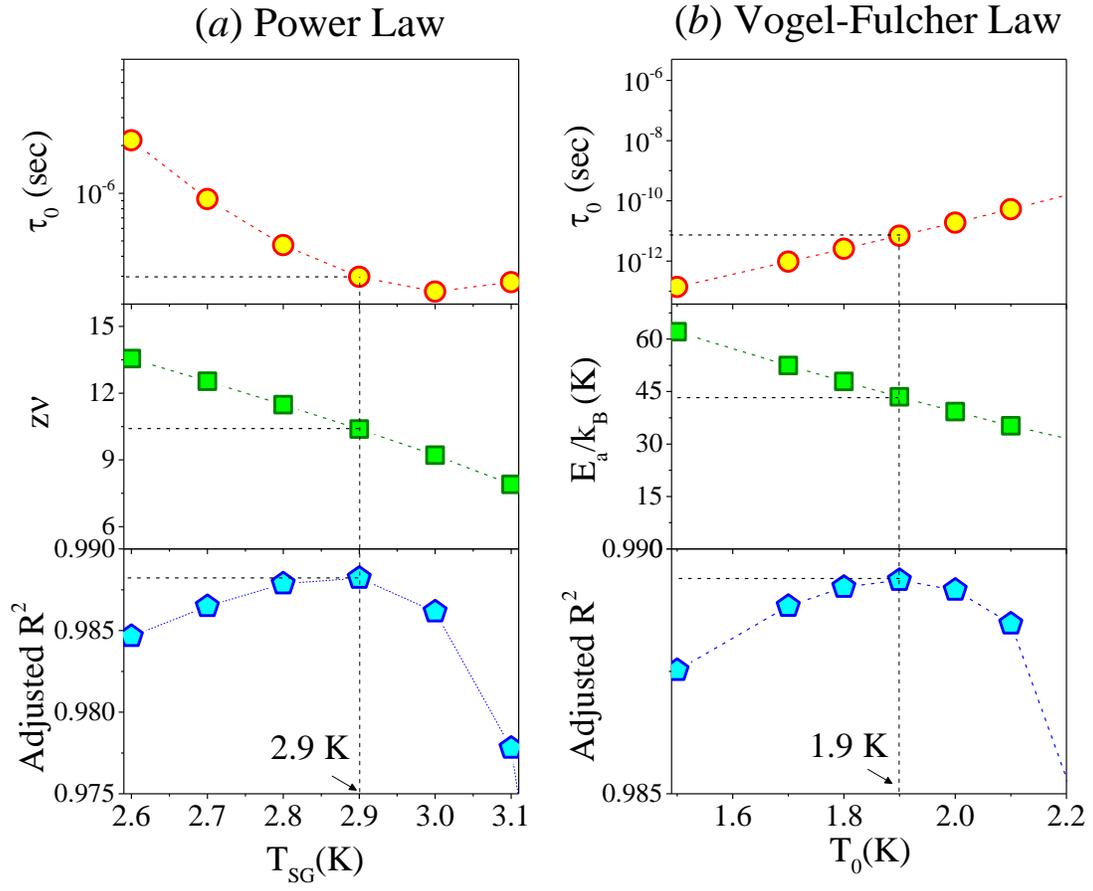

(*a*) Power Law        (*b*) Vogel-Fulcher Law